\documentclass[10pt,twocolumn,english,aps,pra]{revtex4-2}
\usepackage[T1]{fontenc}
\usepackage[utf8]{inputenc}
\usepackage{amsmath}
\usepackage{amsthm}
\usepackage{graphicx}

\makeatletter
\numberwithin{equation}{section}
\numberwithin{figure}{section}

\usepackage{units}
\usepackage{babel}
\usepackage{float}
\usepackage{graphicx}
\usepackage{babel}
\usepackage{enumitem}

\usepackage{tikz}
\usepackage{vmargin}
\setmargins{1.5cm}       % left margin
{0.5cm}                        % top margin
{17.5cm}                      % text width
{25.42cm}                    % text height
{14pt}                           % head height
{1cm}                           % head sep
{0pt}                             % foot height
{0cm}                           % foot skip
\usepackage{amsmath}
\usepackage{amssymb}
\usepackage{cancel}
\usepackage[nodayofweek,yyyymmdd]{datetime}
\usepackage{mdframed}
\definecolor{green}{RGB}{0,128,0}
\addto\captionsenglish{}

\makeatother

\usepackage{babel}
\begin{document}
\global\long\def\blue#1{\textcolor{blue}{#1}}%
\global\long\def\red#1{\textcolor{red}{#1}}%
\global\long\def\green#1{\textcolor{green}{#1}}%
\global\long\def\purple#1{\textcolor{purple}{#1}}%
\global\long\def\orange#1{\textcolor{orange}{#1}}%

\global\long\def\it#1{\textit{\textrm{#1}}}%
\global\long\def\un#1{\underline{\textrm{#1}}}%
\global\long\def\br#1{\left( #1 \right)}%
\global\long\def\sqbr#1{\left[ #1 \right]}%
\global\long\def\curbr#1{\left\{  #1 \right\}  }%
\global\long\def\braket#1{\langle#1 \rangle}%
\global\long\def\bra#1{\langle#1 \vert}%
\global\long\def\ket#1{\vert#1 \rangle}%
\global\long\def\abs#1{\left|#1\right|}%
\global\long\def\mb#1{\mathbf{#1}}%
\global\long\def\doublebraket#1{\langle\langle#1 \rangle\rangle}%

\title{Understanding the Energy Gap Law under Vibrational Strong Coupling}
\author{Yong Rui Poh, Sindhana Pannir-Sivajothi, Joel Yuen-Zhou}
\email{joelyuen@ucsd.edu}

\affiliation{\emph{Department of Chemistry and Biochemistry, University of California
San Diego, La Jolla, California 92093, USA}}
\date{January 5, 2023}
\begin{abstract}
The rate of non-radiative decay between two molecular electronic states
is succinctly described by the energy gap law, which suggests an approximately-exponential
dependence of the rate on the electronic energy gap. Here, we inquire
whether this rate is modified under vibrational strong coupling, a
regime whereby the molecular vibrations are strongly coupled to an
infrared cavity. We show that, under most conditions, the collective
light-matter coupling strength is not large enough to counter the
entropic penalty involved with using the polariton modes, so the energy
gap law remains unchanged. This effect (or the lack thereof) may be
reversed with deep strong light-matter couplings or large detunings,
both of which increase the upper polariton frequency. Finally, we
demonstrate how vibrational polariton condensates mitigate the entropy
problem by providing large occupation numbers in the polariton modes.
\end{abstract}
\maketitle

\section*{Introduction}

When molecules are placed inside an optical cavity, they may interact
strongly with the quantised radiation mode to form light-matter hybrid
states called polaritons \citep{Lidzey1998,Long2015,Shalabney2015,Galego2015,Feist2018}.
In particular, this phenomenon is most significant when energy cycles
between the molecular transitions and the photon mode at a rate faster
than the decay of each individual component. Molecular polaritons
have different frequencies and potential energy surfaces than their
pure-matter counterparts. As such, they offer an interesting avenue
for controlling chemical properties, often through parameters directly
related to light-matter interactions (like cavity frequencies and
light-matter coupling strengths) \citep{Ebbesen2016}. Of particular
interest to this paper are vibrational polaritons, formed by strong
coupling between the high-frequency vibrational modes of molecules
and an infrared optical cavity that houses these molecules \citep{Long2015,Shalabney2015}.
This effect, also known as vibrational strong coupling (VSC), has
been experimentally shown to modify reaction pathways and achieve
chemoselectivity \citep{Thomas2016,Pang2020,GarciaVidal2021,Thomas2019,Hirai2020}.
VSC is an ensemble effect: a large number of molecules, typically
of the order of $10^{10}$, must collectively couple to a cavity mode
to generate an appreciable collective light-matter coupling strength.
The result is the formation of two polariton modes, which provide
control over chemical properties and are desirable, alongside a large
number of dark modes that behave effectively as uncoupled molecules
in the absence of disorder \citep{CamposGonzalezAngulo2020} (with
disorder, dark modes might behave differently from uncoupled molecular
excitations, although the extent to which this occurs is still a subject
of current exploration \citep{Botzung2020,Scholes2020,Du2022}). The
latter is a source of concern; due to their sheer numbers, these non-photonic
dark modes have dominant control over chemical properties of the system,
potentially undoing any benefit created by the polariton modes. In
principle, there exists a specific and reasonable parameter range
that allows some unique properties of vibrational polariton modes
to outweigh the entropic cost from using them \citep{CamposGonzalezAngulo2019};
yet, in practice, these molecular parameter requirements (such as
ultralow reorganisation energies) have not been experimentally found.
Meanwhile, more recently, polariton condensates have been theoretically
shown as an alternative avenue for overcoming the entropic penalty
associated with polariton modes \citep{PannirSivajothi2022,Cortese2017,Phuc2022}.

\begin{figure*}
\includegraphics[width=2\columnwidth]{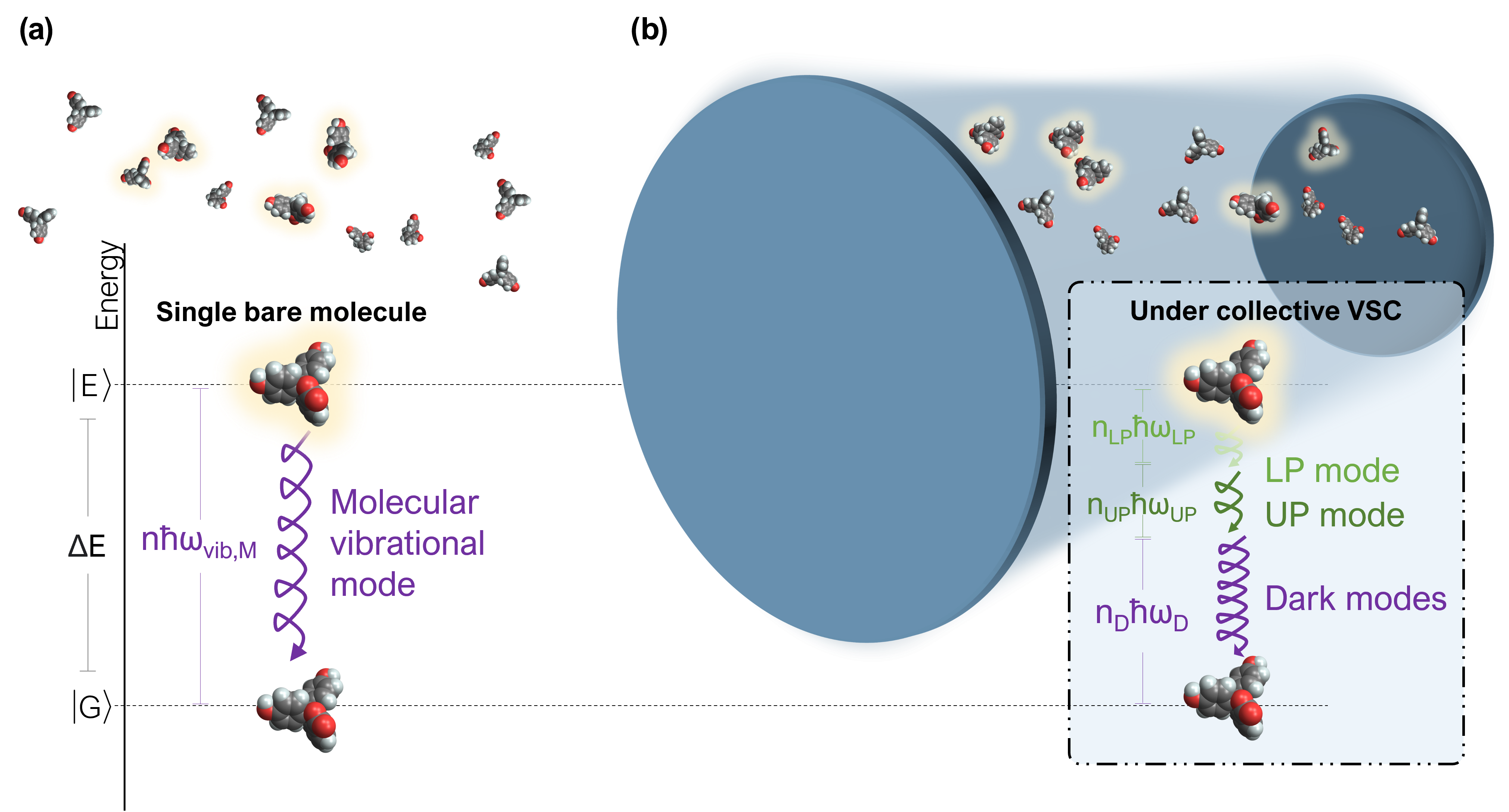}\caption{\label{fig:main}Non-radiative decay of a molecule from a higher electronic
state $\protect\ket{\text{E}}$ to a lower electronic state $\protect\ket{\text{G}}$.
(a) This process may, in its simplest form, be modelled as a non-linear
conversion of energy from electronic energy ($\Delta E$) in state
$\protect\ket{\text{E}}$ to vibrational quanta ($n\hbar\omega_{\text{vib},M}$)
in state $\protect\ket{\text{G}}$ along the vibrational mode of maximum
frequency $\omega_{\text{vib},M}$. (b) Under collective VSC in an
infrared cavity, the same electronic energy ($\Delta E$) may be redistributed
among polariton quanta ($n_{P}\hbar\omega_{P}$, $P=\text{LP},\text{UP}$)
and dark mode quanta ($n_{\text{D}}\hbar\omega_{\text{D}}$), the
former of which is useful because the UP mode has a higher frequency
($\omega_{\text{UP}}$) than the dark modes ($\omega_{\text{D}}=\omega_{\text{vib},M}$)
and can enhance the rate of non-radiative decay. However, under most
circumstances, this advantage is not realised because the large number
of dark modes makes it entropically unfavourable to decay through
the polariton modes.}
\end{figure*}

One chemical property that may benefit from VSC is the non-radiative
decay between electronic states of an excited molecule (Fig. \ref{fig:main}a).
The decay rate is concisely and elegantly described by the energy
gap law \citep{Englman1970,Fischer1970}, which was first derived
by Englman and Jortner in 1970 \citep{Englman1970}. There, they applied
Fermi's golden rule to approximate the transfer rate between two displaced
harmonic oscillators, representing the potential energy surfaces of
two electronic states ($\ket{\text{G}}$ and $\ket{\text{E}}$), through
a diabatic coupling term of amplitude $\abs{J_{\text{GE}}}$ that
is treated perturbatively. More specifically, they considered the
Hamiltonian 
\begin{align}
H_{0} & =\sum_{m}\hbar\omega_{\text{vib},m}\hat{b}_{\text{G},m}^{\dagger}\hat{b}_{\text{G},m}\ket{\text{G}}\bra{\text{G}}\nonumber \\
 & \quad+\br{\sum_{m}\hbar\omega_{\text{vib},m}\hat{b}_{\text{E},m}^{\dagger}\hat{b}_{\text{E},m}+\Delta E}\ket{\text{E}}\bra{\text{E}},
\end{align}
and a perturbative coupling term $V_{\text{trs}}=J_{\text{GE}}\br{\ket{\text{E}}\bra{\text{G}}+\ket{\text{G}}\bra{\text{E}}}$,
where $\Delta E$ is the electronic energy gap, $b_{x,m}^{\dagger}$
($b_{x,m}$) is the creation (annihilation) operator of the $m$-th
vibrational mode in electronic state $x$ with frequency $\omega_{\text{vib},m}$,
and $b_{\text{E},m}=b_{\text{G},m}-\sqrt{S_{m}}$, with $S_{m}$ being
the Huang-Rhys factor representing the displacement of the two potential
energy surfaces along mode $m$. In the low temperature limit, the
non-radiative decay rate is approximately
\begin{align}
W_{\text{bare}} & \approx\frac{\abs{J_{\text{GE}}}^{2}}{\hbar}\sqrt{\frac{2\pi}{\hbar\omega_{\text{vib},M}\Delta E}}\nonumber \\
 & \quad\times e^{-\sum_{m}S_{m}}\exp\br{-\gamma\frac{\Delta E}{\hbar\omega_{\text{vib},M}}},\label{eq:W_bare}
\end{align}
with $\gamma=\ln\frac{\Delta E}{S_{M}\hbar\omega_{\text{vib},M}}-1$.
Here, $\omega_{\text{vib},M}$ is the maximum vibrational frequency
of the molecule, $S_{M}=\sum_{m\in\curbr{\mathcal{M}}}S_{m}$ is the
sum of the Huang-Rhys factors for the set of vibrational modes $\curbr{\mathcal{M}}$
with frequencies near $\omega_{\text{vib},M}$, and the subscript
``bare'' indicates that the rate expression is computed for a molecule
outside the cavity. There are theoretical foundations in grouping
multiple vibrational modes into a single, effective vibrational mode
with contributions from the highest frequency modes \citep{Englman1970}.
The inherent assumption in this expression is that the electronic
energy is most likely lost through a number of high-frequency vibrations
rather than many more low-frequency ones, valid when the total reorganisation
energy $\sum_{m}S_{m}\hbar\omega_{m}$ of the high-frequency modes
is much larger than that of the low-frequency modes \citep{Jang2021}.
An analytical generalisation of this model to account for low-frequency
modes and finite temperatures has been provided by Jang \citep{Jang2021}.
In that work, Jang derives an improved rate expression that retains
the same qualitative phenomena as Eq. (\ref{eq:W_bare}) but with
improved accuracy as compared to numerical results. Since Eq. (\ref{eq:W_bare})
is often fitted into experimental data to obtain $J_{\text{GE}}$,
one can potentially achieve more accurate estimates of $J_{\text{GE}}$
with Jang's improved model. Going back to Eq. (\ref{eq:W_bare}),
in most cases, $\gamma$ may be regarded as a constant, giving an
approximately-exponential dependence of the decay rate on the energy
gap $\Delta E$ (hence the name). Overall, this model has found numerous
applications in molecular spectroscopy \citep{Chynwat1995,Wilson2001,Caspar1982,Martin1975}
and optoelectronics \citep{TuongLy2017,Wang2017,ColladoFregoso2019,Wei2020,Zhang2021,Friedman2021},
in particular for describing the quantum yields of radiative processes.

In this paper, we demonstrate how, under VSC, a large proportion of
non-radiative decay occurs through the dark modes, which was expected
due to their large numbers (Fig. \ref{fig:main}b). At the same time,
decay through the higher-frequency polariton channel, if significant,
can reduce the effective energy gap for dark mode decay, thereby increasing
the overall decay rate. These two effects work against each other
and the polaritonic one dominates only under extreme conditions such
as deep strong couplings and large detunings; otherwise, the rate
of non-radiative decay, being dominated by the dark modes, takes a
value similar to the bare one outside of the cavity. Finally, we investigate
how this entropic problem may be mitigated by the use of polariton
condensates.

\section*{Results and Discussions}

\subsection*{The model system}

We consider a model system of $N$ identical molecules $i=1,\cdots,N$,
each with two electronic states ($\ket{\text{E}_{i}}$ and $\ket{\text{G}_{i}}$)
and a set of molecular vibrational modes. The collective effects of
these vibrational modes are represented by a single mode at the maximum
frequency $\omega_{\text{vib}}$ with a collective Huang-Rhys factor
$S$; this is consistent with the approach taken by Englman and Jortner
\citep{Englman1970}. Along this effective vibrational coordinate,
the potential energy curves for both electronic states are modelled
as a pair of displaced harmonic oscillators, each with the same frequency.
Finally, all $N$ molecules are strongly coupled to a lossless cavity
mode of frequency $\omega_{\text{ph}}$ that is in (or close to) resonance
with the effective vibrational mode. To zeroth-order in the diabatic
coupling $J_{\text{GE}}$, the Hamiltonian reads
\begin{align}
\hat{H}_{0} & =\hbar\omega_{\text{ph}}\hat{a}_{\text{ph}}^{\dagger}\hat{a}_{\text{ph}}\nonumber \\
 & \quad+\sum_{i=1}^{N}\hbar\omega_{\text{vib}}\hat{b}_{\text{G},i}^{\dagger}\hat{b}_{\text{G},i}\ket{\text{G}_{i}}\bra{\text{G}_{i}}\nonumber \\
 & \quad+\sum_{i=1}^{N}\br{\hbar\omega_{\text{vib}}\hat{b}_{\text{E},i}^{\dagger}\hat{b}_{\text{E},i}+\Delta E}\ket{\text{E}_{i}}\bra{\text{E}_{i}}\nonumber \\
 & \quad+\sum_{i=1}^{N}\hbar g\br{\hat{b}_{\text{G},i}^{\dagger}\hat{a}_{\text{ph}}+\hat{a}_{\text{ph}}^{\dagger}\hat{b}_{\text{G},i}}\ket{\text{G}_{i}}\bra{\text{G}_{i}}\nonumber \\
 & \quad+\sum_{i=1}^{N}\hbar g\br{\hat{b}_{\text{E},i}^{\dagger}\hat{a}_{\text{ph}}+\hat{a}_{\text{ph}}^{\dagger}\hat{b}_{\text{E},i}}\ket{\text{E}_{i}}\bra{\text{E}_{i}},
\end{align}
where $\hat{a}_{\text{ph}}^{\dagger}\br{\hat{a}_{\text{ph}}}$ is
the creation (annihilation) operator of the photon mode, $\hat{b}_{x,i}^{\dagger}\br{\hat{b}_{x,i}}$
is the creation (annihilation) operator of the $i$-th molecule's
vibrational mode ($x=\text{G},\text{E}$) with $\hat{b}_{\text{E},i}=\hat{b}_{\text{G},i}-\sqrt{S}$
for all $i$, $\Delta E$ is the energy gap between the two electronic
states of each molecule and $g$ is the single-molecule light-matter
coupling strength. Here, we have made the rotating wave approximation
(RWA) and assumed that $g$ is the same for both electronic states.
Also, zero-point energies of the vibrational states have been omitted
since they only contribute constants to the final energies.

To model the non-radiative decay rate of a single molecule $c$ from
the $\ket{\text{E}_{c}}$ electronic state to the $\ket{\text{G}_{c}}$
electronic state, we introduce a diabatic coupling term into the Hamiltonian,
\begin{align}
\hat{V}_{\text{trs},c} & =J_{\text{GE}}\br{\ket{\text{E}_{c}}\bra{\text{G}_{c}}+\ket{\text{G}_{c}}\bra{\text{E}_{c}}},
\end{align}
where $J_{\text{GE}}$ is the corresponding amplitude. Then, following
the same procedure as Englman and Jortner \citep{Englman1970}, we
may compute the non-radiative decay rate of molecule $c$ by Fermi's
golden rule, which assumes $\hat{V}_{\text{trs},c}$ to be a perturbation
with respect to $\hat{H}_{0}$. Despite the presence of light-matter
couplings in $\hat{H}_{0}$, this assumption remains valid since non-radiative
decay effectively couples a single electronic excitation to a large
number of vibrational excitations $n\approx\frac{\Delta E}{\hbar\omega_{\text{vib}}}\gg1$,
all within the same molecule. As such, the nonlinearity of this process
makes it slower than the molecule's interactions with the cavity mode,
which is linear. More quantitatively, non-radiative decay is characterised
by the decay amplitude multiplied by the Franck-Condon overlap between
the initial and final vibrational states of the decaying molecule;
in the low-temperature limit, this takes the form of
\begin{align}
J_{\text{GE}}\sqrt{e^{-S}\frac{S^{n}}{n!}} & \br{\approx10^{-9}\hbar\omega_{\text{vib}}},
\end{align}
which is two orders of magnitude smaller than the decaying molecule's
light-matter coupling strength $\hbar g\br{\approx10^{-7}\hbar\omega_{\text{vib}}}$
if we consider $J_{\text{GE}}\approx0.4\hbar\omega_{\text{vib}}$,
$g\sqrt{N}\approx0.01\omega_{\text{vib}}$, $S\approx0.1$, $\Delta E\approx10\hbar\omega_{\text{vib}}$
and $N\approx10^{10}$, conditions typical of $S_{1}\rightarrow S_{0}$
transitions of aromatic hydrocarbons \citep{Byrne1965,Chynwat1995}
under collective VSC \citep{Hirai2020}.

To find the initial and final eigenstates of $\hat{H}_{0}$, we may,
without loss of generality, focus on the decay of molecule 1 from
$\ket{\text{E}}$ to $\ket{\text{G}}$ while keeping the remaining
$N-1$ molecules in $\ket{\text{G}}$ (i.e. pick $c=1$). Then, the
initial and final electronic eigenstates of $\hat{H}_{0}$ may be
written collectively as $\ket{\text{E},\text{G},\cdots,\text{G}}$
and $\ket{\text{G},\text{G},\cdots,\text{G}}$, where we have listed
the electronic state of each molecule in increasing order of its index
$i$. Each of the two electronic states above comprises $N$ vibrational
modes coupled to a single cavity mode. Therefore, all that remains
is finding the normal modes of $\hat{H}_{0}$ within these two electronic
subspaces. Working first in the subspace of the initial electronic
state, we transform the $N$ vibrational modes into a single bright
(B) mode,
\begin{align}
\hat{b}_{\text{B}} & =\frac{1}{\sqrt{N}}\br{\hat{b}_{\text{E},1}+\sum_{i=2}^{N}\hat{b}_{\text{G},i}},
\end{align}
with the correct symmetry to interact with light and $N-1$ dark (D)
modes,
\begin{align}
\hat{b}_{\text{D},k} & =C_{k,1}\hat{b}_{\text{E},1}+\sum_{i=2}^{N}C_{k,i}\hat{b}_{\text{G},i},
\end{align}
with $2\le k\le N$, that do not couple to light. Note that the constants
$\curbr{C_{k,i}}$ ($1\le i\le N$) are chosen such that the dark
modes are orthogonal to the bright mode, i.e. $\sum_{i=1}^{N}C_{k,i}=0$,
and to each other, i.e. $\sum_{i=1}^{N}C_{j,i}^{*}C_{k,i}=\delta_{jk}$.
In this basis, the dark modes are already diagonal while the bright
and photon modes mix to give the upper polariton (UP) $\hat{b}_{\text{UP}}$
and lower polariton (LP) $\hat{b}_{\text{LP}}$ modes,
\begin{align}
\hat{b}_{\text{LP}} & =-\sin\br{\theta}\hat{a}_{\text{ph}}+\cos\br{\theta}\hat{b}_{\text{B}},\nonumber \\
\hat{b}_{\text{UP}} & =\cos\br{\theta}\hat{a}_{\text{ph}}+\sin\br{\theta}\hat{b}_{\text{B}},
\end{align}
with a mixing angle of
\begin{align}
\theta & =\tan^{-1}\br{\frac{\Omega-\Delta}{2\sqrt{N}g}},
\end{align}
where $\Omega=\sqrt{\Delta^{2}+4g^{2}N}$ is the Rabi splitting, $\Delta=\omega_{\text{ph}}-\omega_{\text{vib}}$
is the detuning and $g\sqrt{N}$ is the collective light-matter coupling
strength. Also, the mode frequencies are
\begin{align}
\omega_{\text{LP}} & =\omega_{\text{vib}}+\frac{\Delta-\Omega}{2},\nonumber \\
\omega_{\text{UP}} & =\omega_{\text{vib}}+\frac{\Delta+\Omega}{2},\nonumber \\
\omega_{\text{D},k} & =\omega_{\text{vib}} &  & \text{ for all }k.
\end{align}
We may follow the same steps to define the bright, dark and polariton
modes $\hat{b}_{\text{B}}'$, $\curbr{\hat{b}_{\text{D},k}'}$, $\hat{b}_{\text{LP}}'$
and $\hat{b}_{\text{UP}}'$ in the final electronic state subspace,
where we shall prime quantities belonging to the final state and unprime
those for the initial state. Note that, since vibrational modes have
the same light-matter coupling in both electronic states, the primed
modes have the same frequencies as the unprimed ones, i.e. $\omega_{\text{LP}'}=\omega_{\text{LP}}$,
$\omega_{\text{UP}'}=\omega_{\text{UP}}$ and $\omega_{\text{D}',k}=\omega_{\text{D},k}=\omega_{\text{vib}}$.
Notice also how, during the decay of molecule 1 from $\ket{\text{E}}$
to $\ket{\text{G}}$, all $N+1$ modes change,
\begin{gather}
\text{LP}+\text{UP}+\sum_{k=1}^{N-1}\text{D}_{k}\longrightarrow\text{LP}'+\text{UP}'+\sum_{k=1}^{N-1}\text{D}_{k}'.
\end{gather}
However, by writing the dark modes in a highly localised basis \citep{CamposGonzalezAngulo2019,Strashko2016,Heller2018},
contributions from molecule 1 are completely localised onto one dark
mode $\text{D}_{\text{loc}}$, such that we may reduce the number
of reacting modes to three,
\begin{gather}
\text{LP}+\text{UP}+\text{D}_{\text{loc}}\longrightarrow\text{LP}'+\text{UP}'+\text{D}_{\text{loc}}'.
\end{gather}
Specifically, in this basis, the relevant dark mode operators are
\begin{align}
\hat{b}_{\text{D},\text{loc}} & =\sqrt{\frac{N-1}{N}}\hat{b}_{\text{E},1}-\sqrt{\frac{1}{N\br{N-1}}}\sum_{i=2}^{N}\hat{b}_{\text{G},i},\\
\hat{b}_{\text{D},\text{loc}}' & =\sqrt{\frac{N-1}{N}}\hat{b}_{\text{G},1}-\sqrt{\frac{1}{N\br{N-1}}}\sum_{i=2}^{N}\hat{b}_{\text{G},i}.
\end{align}
More details can be found in Supplementary Note 1. 

\subsection*{Evaluating the golden rule expression}

Since the energy gap law is most widely applied in the low temperature
limit, the same assumption is made here whereby all $N$ molecules
are in the ground vibrational state of their respective electronic
states, given by $\ket{0^{\br{\text{LP}}},0^{\br{\text{UP}}},0^{\br{\text{D}}}}$.
Here and hereafter, we will label eigenstates of the zeroth-order
Hamiltonian $\hat{H}_{0}$ within the electronic subspace by three
numbers representing the occupancy numbers in the LP, UP and localised
D modes respectively. By applying first-order time-dependent perturbation
theory, we arrive at the expression
\begin{align}
 & W_{\text{VSC}}\nonumber \\
 & =\frac{2\pi}{\hbar}\abs{J_{\text{GE}}}^{2}\sum_{n_{\text{LP}},n_{\text{UP}},n_{\text{D}}=0}^{\infty}\nonumber \\
 & \quad\delta\br{\Delta E-n_{\text{LP}}\hbar\omega_{\text{LP}}-n_{\text{UP}}\hbar\omega_{\text{UP}}-n_{\text{D}}\hbar\omega_{\text{vib}}}\nonumber \\
 & \quad\times\abs{\braket{n_{\text{LP}}^{\br{\text{LP}'}},n_{\text{UP}}^{\br{\text{UP}'}},n_{\text{D}}^{\br{\text{D}'}}\vert0^{\br{\text{LP}}},0^{\br{\text{UP}}},0^{\br{\text{D}}}}}^{2},
\end{align}
where $W_{\text{VSC}}$ is the non-radiative decay rate under VSC
and $\delta$ is the Dirac delta function. It may be shown that the
potential energy surfaces of the initial and final electronic states
along the LP, UP and D modes behave as displaced harmonic oscillators
too, with effective Huang-Rhys factors of $S_{\text{LP}}=\frac{S}{N}\cos^{2}\theta$,
$S_{\text{UP}}=\frac{S}{N}\sin^{2}\theta$ and $S_{\text{D}}=\frac{S}{N}\br{N-1}$
respectively (see Supplementary Note 2). By expressing the Dirac delta
function in its Fourier form and evaluating the Franck-Condon overlap
of displaced harmonic oscillators, we get
\begin{align}
 & W_{\text{VSC}}\nonumber \\
 & =\frac{\abs{J_{\text{GE}}}^{2}}{\hbar\br{\hbar\omega_{\text{vib}}}}e^{-S}\sum_{n_{\text{LP}},n_{\text{UP}},n_{\text{D}}=0}^{\infty}\int_{-\infty}^{\infty}dt\,\nonumber \\
 & \quad\exp\br{i\frac{\Delta E}{\hbar\omega_{\text{vib}}}t-in_{\text{LP}}\bar{\omega}_{\text{LP}}t-in_{\text{UP}}\bar{\omega}_{\text{UP}}t-in_{\text{D}}t}\nonumber \\
 & \quad\times\frac{\br{S_{\text{LP}}}^{n_{\text{LP}}}}{n_{\text{LP}}!}\frac{\br{S_{\text{UP}}}^{n_{\text{UP}}}}{n_{\text{UP}}!}\frac{\br{S_{\text{D}}}^{n_{\text{D}}}}{n_{\text{D}}!},\label{eq:W_VSC-general}
\end{align}
where $\bar{\omega}_{P}=\omega_{P}/\omega_{\text{vib}}$ is the frequency
of the polariton modes $\text{P}=\text{LP},\text{UP}$ relative to
that of the dark mode (as well as the bare molecule) and $t$ is dimensionless.
Note that in the original work by Englman and Jortner \citep{Englman1970},
the integral in Eq. (\ref{eq:W_VSC-general}) was evaluated via two
approaches: (1) by performing a short time expansion, and (2) using
the saddle point approximation. The first approach is valid when the
vibronic couplings (characterised by the Huang-Rhys factors) are large,
such that non-radiative relaxation occurs predominantly through a
thermally activated pathway. Indeed, the resulting rate expression
has an exponential dependence on the activation energy and has been
used to describe electron transfer rates through the celebrated Marcus
\citep{Marcus1956} and Marcus-Levich-Jortner \citep{Marcus1964,Levich1966,Jortner1976}
theories. Applications of these theories to VSC have been explored
by our group \citep{CamposGonzalezAngulo2019} and the Phuc and Ishizaki
groups \citep{Phuc2020} and will not be discussed here. In contrast,
the second approach is useful when the vibronic couplings are weak,
such that nuclear tunneling is the main non-radiative decay pathway.
This approach is more applicable to intersystem crossing and internal
conversions -- processes which are experimentally modelled by the
energy gap law \citep{Chynwat1995,Wilson2001,Caspar1982,Martin1975,TuongLy2017,Wang2017,ColladoFregoso2019,Wei2020,Zhang2021,Friedman2021}
-- and forms the focus of this paper. These differences have been
detailed by Nitzan \citep{Nitzan2006-EGL} and a useful guide on the
saddle point method may be found in Morse and Feshbach \citep{Morse1953}.
More recently, an evaluation of such rate expressions through a path-integral
approach \citep{Kessing2022} and the general case of VSC-mediated
resonance energy transfer \citep{Cao2022} have also been considered
by the Cao group.

Returning to Eq. (\ref{eq:W_VSC-general}), one way to move forward
is to bring all three summations into the integral and perform a single
saddle point approximation (see Supplementary Note 3) to obtain 
\begin{align}
W_{\text{VSC}}^{\text{one}} & \approx\frac{\abs{J_{\text{GE}}}^{2}}{\hbar\br{\hbar\omega_{\text{vib}}}}e^{-S}\sqrt{\frac{2\pi}{-f''\br{\tau_{0}}}}e^{-f\br{\tau_{0}}},\label{eq:W_VSC^one}
\end{align}
where $f\br{\tau}=-\frac{\Delta E}{\hbar\omega_{\text{vib}}}\tau-S_{\text{LP}}e^{-\bar{\omega}_{\text{LP}}\tau}-S_{\text{UP}}e^{-\bar{\omega}_{\text{UP}}\tau}-S_{\text{D}}e^{-\tau}$
and $\tau=\tau_{0}$ is the extremum of $\text{Re }f\br{\tau}$. This
approach has the effect of taking the final vibrational states of
all three modes to the continuum limit and is valid for $\frac{\Delta E}{\hbar\omega_{\text{vib}}}\gg1$.
Finding $\tau_{0}$ is, however, challenging and involves solving
a transcendental equation. Here, we seek an analytical solution for
$W_{\text{VSC}}^{\text{one}}$ by setting the detuning $\Delta$ to
zero, expanding $\tau_{0}$ in powers of $\frac{g\sqrt{N}}{\omega_{\text{vib}}}$
and keeping the leading term to obtain
\begin{align}
W_{\text{VSC}}^{\text{one}} & \approx\frac{\abs{J_{\text{GE}}}^{2}}{\hbar}e^{-S}\sqrt{\frac{2\pi}{\hbar\omega_{\text{vib}}\Delta E\phi}}\exp\sqbr{-\Gamma\frac{\Delta E}{\hbar\omega_{\text{vib}}}},
\end{align}
with 
\begin{align}
\Gamma & =\gamma-\br{\frac{g}{\omega_{\text{vib}}}}^{2}\sqbr{\frac{1}{2}\br{\ln\frac{\Delta E}{S\hbar\omega_{\text{vib}}}}^{2}}\nonumber \\
 & \quad+\mathcal{O}\sqbr{\br{\frac{g\sqrt{N}}{\omega_{\text{vib}}}}^{4}},\\
\phi & =1+\br{\frac{g}{\omega_{\text{vib}}}}^{2}\br{\ln\frac{\Delta E}{S\hbar\omega_{\text{vib}}}+1}\nonumber \\
 & \quad+\mathcal{O}\sqbr{\br{\frac{g\sqrt{N}}{\omega_{\text{vib}}}}^{4}}.
\end{align}
where $\gamma=\ln\frac{\Delta E}{S\hbar\omega_{\text{vib}}}-1$. Comparing
this result with Eq. (\ref{eq:W_bare}), we notice that the first
terms give the bare molecule rate while subsequent terms serve as
corrections due to coupling to the cavity. Surprisingly, the first
correction terms do not depend on $N$, a result that has been corroborated
by Yang and Cao \citep{Yang2021} and more recently by Kansanen \citep{Kansanen2022}.
Therefore, for some constant and small $g\sqrt{N}$ (relative to $\omega_{\text{vib}}$),
we expect VSC to enhance the decay rate at small $N$, since this
implies having a system with larger $g$. Physically, this implies
that the cavity's coupling to the single decaying molecule dominates
the relaxation dynamics, and further interactions between the $N-1$
non-decaying molecules and the single decaying one, through the cavity,
appear only as higher-order processes. This effect, which is further
explained by Du and co-workers \citep{Du2022-PGH}, is essentially
the message from the polariton ``large $N$ problem'' \citep{MartinezMartinez2019}
(more to follow).

We now consider the case whereby, starting from Eq. (\ref{eq:W_VSC-general}),
only the sum over final dark states is brought into the integral.
Many saddle point approximations are now required, one for every $n_{\text{LP}}$
and $n_{\text{UP}}$ term, and we obtain
\begin{align}
W_{\text{VSC}}^{\text{many}} & \approx\frac{\abs{J_{\text{GE}}}^{2}}{\hbar}e^{-S}\sum_{n_{\text{LP}},n_{\text{UP}}=0}^{\infty}w_{\text{VSC}}^{\text{many}}\br{n_{\text{LP}},n_{\text{UP}}}\label{eq:W_VSC^many}
\end{align}
with
\begin{align}
 & w_{\text{VSC}}^{\text{many}}\br{n_{\text{LP}},n_{\text{UP}}}\nonumber \\
 & =\frac{\br{S_{\text{LP}}}^{n_{\text{LP}}}}{n_{\text{LP}}!}\frac{\br{S_{\text{UP}}}^{n_{\text{UP}}}}{n_{\text{UP}}!}\nonumber \\
 & \quad\times\sqrt{\frac{2\pi}{\hbar\omega_{\text{vib}}\Delta\tilde{E}}}\exp\br{-\tilde{\gamma}\frac{\Delta\tilde{E}}{\hbar\omega_{\text{vib}}}},
\end{align}
where $\tilde{\gamma}=\ln\frac{\Delta\tilde{E}}{S_{\text{D}}\hbar\omega_{\text{vib}}}-1$
and $\Delta\tilde{E}=\Delta E-n_{\text{LP}}\hbar\omega_{\text{LP}}-n_{\text{UP}}\hbar\omega_{\text{UP}}$
is the effective energy gap for dark mode decay after subtracting
any energy distributed through the LP and UP modes. This result is
valid in the limit of $\frac{\Delta\tilde{E}}{\hbar\omega_{\text{vib}}}\gg1$
for every $\br{n_{\text{LP}},n_{\text{UP}}}$ term (see Supplementary
Note 4); effectively, this approach assumes a continuum of final dark
states with discrete final polariton states or, equivalently, that
most of the energy is being distributed through the dark mode. With
this alternative, more intuitive approach, the polariton modes' contribution
to the decay rate may be separately identified. For instance, the
$\br{n_{\text{LP}},n_{\text{UP}}}=\br{0,0}$ term in Eq. (\ref{eq:W_VSC^many})
reduces to the single molecule decay rate $W_{\text{bare}}$ in the
large $N$ limit. This is the case whereby all of the energy is transferred
through the dark mode. As such, terms of $n_{\text{LP}},n_{\text{UP}}>0$
signify the polaritons' contributions to the decay rate and are generally
small. For instance, $w_{\text{VSC}}^{\text{many}}\br{0,1}$ is approximately
\begin{align}
w_{\text{VSC}}^{\text{many}}\br{0,1} & \approx S_{\text{UP}}e^{\gamma}w_{\text{VSC}}^{\text{many}}\br{0,0}.\label{eq:VSC_condition-1}
\end{align}
where $\gamma=\ln\frac{\Delta E}{S_{\text{D}}\hbar\omega_{\text{vib}}}-1$
and we have assumed $\omega_{\text{UP}}\approx\omega_{\text{vib}}$,
which is true under typical conditions for VSC. Note that, near resonance,
$S_{\text{UP}}\simeq\frac{S}{2N}$. Since $N$ is usually large, we
expect $S_{\text{UP}}$ to be small and $w_{\text{VSC}}^{\text{many}}\br{0,1}\ll w_{\text{VSC}}^{\text{many}}\br{0,0}$.
Similar arguments may be made for larger $n_{\text{UP}}$ terms and
for $n_{\text{LP}}$ terms (see Supplementary Note 5). Indeed, decay
through the dark mode dominates the overall non-radiative decay rate.
This is a manifestation of the ``large $N$ problem'' \citep{MartinezMartinez2019}:
while using the UP mode reduces the effective energy gap experienced
by the dark modes, thereby increasing the non-radiative decay rate
through the $e^{\gamma}$ factor in Eq. (\ref{eq:VSC_condition-1}),
this benefit is hampered by the entropic penalty from the polariton
modes each containing only $\approx\frac{1}{2N}$ part of a single
molecule, where $N$ is typically large.

For VSC to be useful, we want the case of $w_{\text{VSC}}^{\text{many}}\br{0,1}>w_{\text{VSC}}^{\text{many}}\br{0,0}$,
that is, when decay through the UP mode becomes significant. Near
resonance, this condition simplifies to
\begin{align}
N & \lesssim\frac{1}{5}\frac{\Delta E}{\hbar\omega_{\text{vib}}}.\label{eq:VSC_condition}
\end{align}
With $W_{\text{VSC}}^{\text{one}}$, we concluded that, for some constant
and small $g\sqrt{N}/\omega_{\text{vib}}$, rate enhancement under
VSC may only be achieved at small $N$. Here, we specify the condition
of small $N$ with Eq. (\ref{eq:VSC_condition}): $N$ must be smaller
than the electronic energy gap relative to the vibrational frequency.
This condition arises because the system only sees the $\frac{1}{2N}$
entropic penalty and not the $g\sqrt{N}$ frequency increase in the
UP mode; this again explains why $W_{\text{VSC}}^{\text{one}}$ had
no $N$ dependence to leading order in $g\sqrt{N}$.

In practice, $\Delta E$ is usually around $7000-20000\mathrm{\,cm^{-1}}$
and $\omega_{\text{vib}}$ is usually chosen to be $3000\mathrm{\,cm^{-1}}$
(corresponding to the C$-$H stretch) \citep{Maciejewski1984,Friedman2021}.
This gives $\frac{\Delta E}{\hbar\omega_{\text{vib}}}$ to be $\approx2-7$.
Consequently, we must reach the $N\rightarrow1$ limit, i.e. single
molecule coupling limit, before we can observe rate enhancement under
VSC. Such couplings may be achieved in nanophotonic cavities \citep{Chikkaraddy2016,Bitton2022};
however, under most circumstances, VSC should have minimal effects
on the energy gap law.

\subsection*{Numerical simulations}

The discussion above may be verified with some numerical simulations.
We consider systems under resonance condition $\Delta=0$ with $g\sqrt{N}=0.01\omega_{\text{vib}}$
and $S=0.1$ over a range of energy gaps $10\le\frac{\Delta E}{\hbar\omega_{\text{vib}}}\le50$
and number of molecules $10\le N\le10^{4}$. These energy gaps are
larger than most chromophores and we expect an even smaller rate enhancement
in usual chromophores; the remaining parameters are typical of non-radiative
decays of aromatic hydrocarbons to the ground state \citep{Byrne1965,Chynwat1995}
under collective VSC \citep{Hirai2020}. Firstly, rates computed using
$W_{\text{VSC}}^{\text{one}}$ and $W_{\text{VSC}}^{\text{many}}$
were within $\sim0.1\%$ of each other, i.e. both ways of applying
saddle point approximations give answers that are numerically similar
(see Supplementary Note 4). Next, from Fig. \ref{fig:VSC_condition},
we verify that rate enhancement is only observed when $N\lesssim\frac{1}{5}\frac{\Delta E}{\hbar\omega_{\text{vib}}}$,
a small parameter range that is difficult to achieve experimentally
and, even so, gives little rate enhancements of $\sim1\%$ higher
rates than the bare molecule case. Finally, by observing the number
of terms added before the sum in $W_{\text{VSC}}^{\text{many}}$ converges,
we conclude that polaritons are indeed responsible for the observed
rate enhancements. In fact, the largest decay rates in Fig. \ref{fig:VSC_condition}
are obtained from close-to-complete decay through the polariton modes.

\begin{figure}
\includegraphics[width=1\columnwidth]{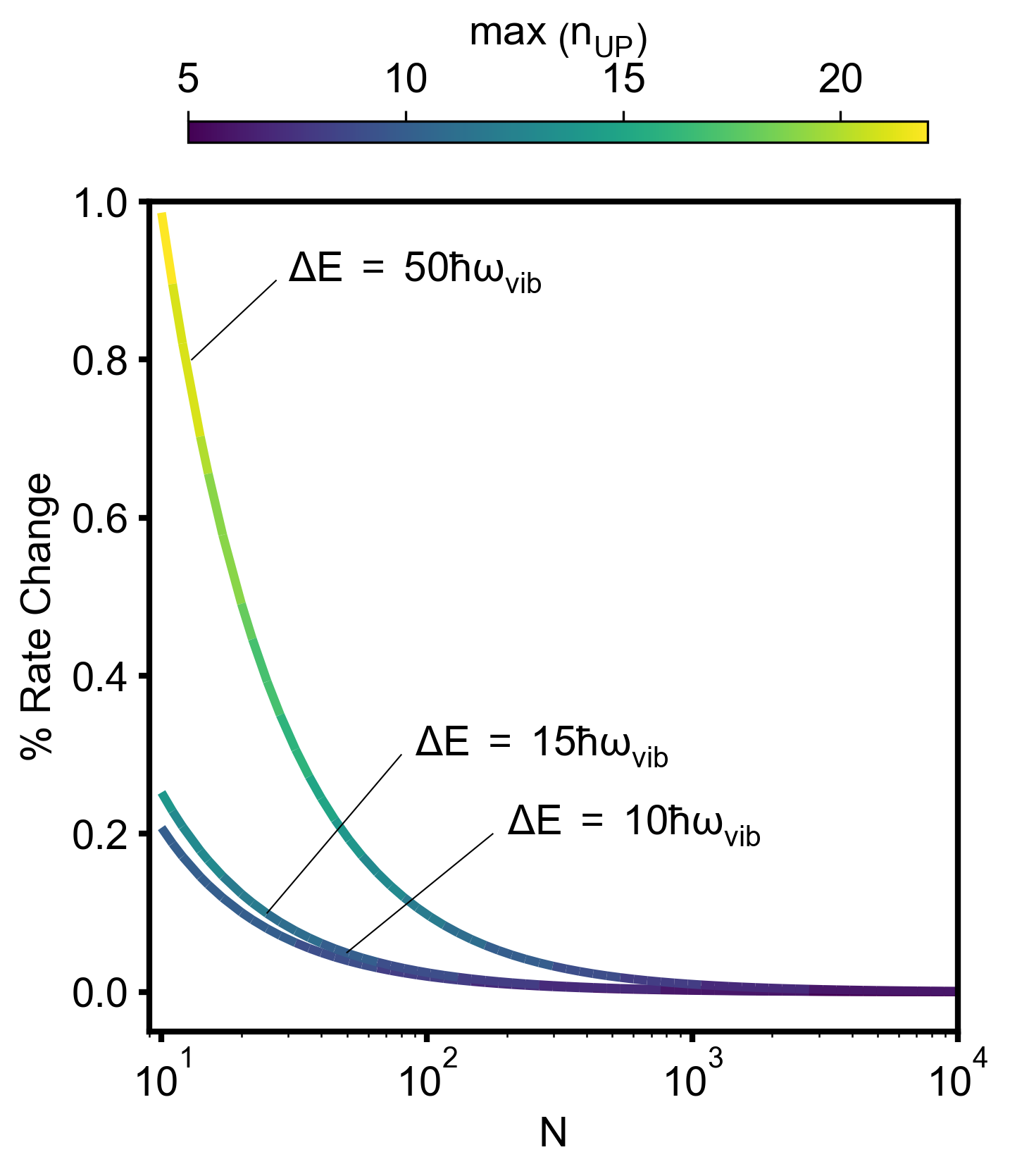}

\caption{\label{fig:VSC_condition}Effect of collective VSC on single-molecule
non-radiative decay rates over ranges of $N$, the number of molecules,
and $\frac{\Delta E}{\hbar\omega_{\text{vib}}}$, the relative electronic
energy gap. Rates inside the cavity were calculated using many saddle
point approximations ($W_{\text{VSC}}^{\text{many}}$; see Eq. (\ref{eq:W_VSC^many}))
and compared with that outside of the cavity ($W_{\text{bare}}$)
to obtain the percentage rate changes, which are small and only substantial
at small $N$. The value $\max\protect\br{n_{\text{UP}}}$ represents
the number of $n_{\text{UP}}$ terms in Eq. (\ref{eq:W_VSC^many})
that needed to be summed before the subsequent term falls below $10^{-15}$.
It increases with increasing rate changes, suggesting that the UP
mode is responsible for the growing decay rates. Similar results were
observed for $\max\protect\br{n_{\text{LP}}}$. All plots were generated
with the following parameters: detuning, $\Delta=0$; collective light-matter
coupling, $g\sqrt{N}=0.01\omega_{\text{vib}}$; bare-molecule Huang-Rhys
factor, $S=0.1$.}
\end{figure}

In the next few sections, we will discuss how we may overcome the
entropic penalty associated with polariton mode decay with (1) higher
UP frequencies and (2) polariton condensates.

\subsection*{Overcoming the entropic penalty with larger UP frequencies}

The UP mode offers a higher frequency decay channel that can potentially
speed up non-radiative decay by reducing the effective energy gap
$\Delta\tilde{E}$ for dark mode decay. However, under most experimental
circumstances, this advantage cannot be exploited to its fullest.
This is because the light-matter coupling strength $g\sqrt{N}$ is
typically small at around $0.01-0.05\omega_{\text{vib}}$, so $\omega_{\text{UP}}\approx\omega_{\text{vib}}$.
What if $\omega_{\text{UP}}$ is appreciably larger than $\omega_{\text{vib}}$?
To answer this, we return to Eq. (\ref{eq:VSC_condition-1}), but,
this time, we do not set $\omega_{\text{UP}}\approx\omega_{\text{vib}}$.
The expression becomes
\begin{align}
w_{\text{VSC}}^{\text{many}}\br{0,1} & \approx S_{\text{UP}}e^{\gamma\bar{\omega}_{\text{UP}}}w_{\text{VSC}}^{\text{many}}\br{0,0},
\end{align}
where $\gamma=\ln\frac{\Delta E}{S_{\text{D}}\hbar\omega_{\text{vib}}}-1$.
Then, decay through the UP mode becomes significant when $w_{\text{VSC}}^{\text{many}}\br{0,1}>w_{\text{VSC}}^{\text{many}}\br{0,0}$,
which simplifies to
\begin{align}
\frac{S_{\text{D}}}{S_{\text{UP}}} & \lesssim\frac{\Delta E}{\hbar\omega_{\text{vib}}}e^{-1+\gamma\br{\bar{\omega}_{\text{UP}}-1}}.\label{eq:VSC_condition-large_UP_freq-1}
\end{align}
Near resonance, we have $S_{\text{UP}}\simeq\frac{S}{2N}$ and $\bar{\omega}_{\text{UP}}\simeq1+\frac{g\sqrt{N}}{\omega_{\text{vib}}}$,
so the condition becomes
\begin{align}
N & \lesssim\frac{1}{5}\frac{\Delta E}{\hbar\omega_{\text{vib}}}e^{\gamma\br{\frac{g\sqrt{N}}{\omega_{\text{vib}}}}}.\label{eq:VSC_condition-large_UP_freq}
\end{align}
which suggests a competition between the UP mode frequency $\bar{\omega}_{\text{UP}}$
improving decay rates and the entropic penalty $\frac{1}{2N}$ reversing
this effect. Here, we want the former to dominate over the latter.

For larger $\omega_{\text{UP}}$ values, the validity of rates computed
using $W_{\text{VSC}}^{\text{one}}$ (Eq. (\ref{eq:W_VSC^one})) cannot
be easily verified. While there are techniques that may overcome this
problem \citep{Ansari2022,Fang2019,Li2022,Kessing2022}, for simplicity,
we choose to rewrite the integral in Eq. (\ref{eq:W_VSC-general})
into another form so that we can continue to apply the saddle point
approximation. This limitation and the alternative taken are detailed
in Supplementary Note 6, and rates evaluated using this new approach
will be labelled as $\bar{\bar{W}}_{\text{VSC}}^{\text{one}}$.

\subsection*{Increasing the upper polariton frequency by deep strong light-matter
couplings}

\begin{figure}
\includegraphics[width=1\columnwidth]{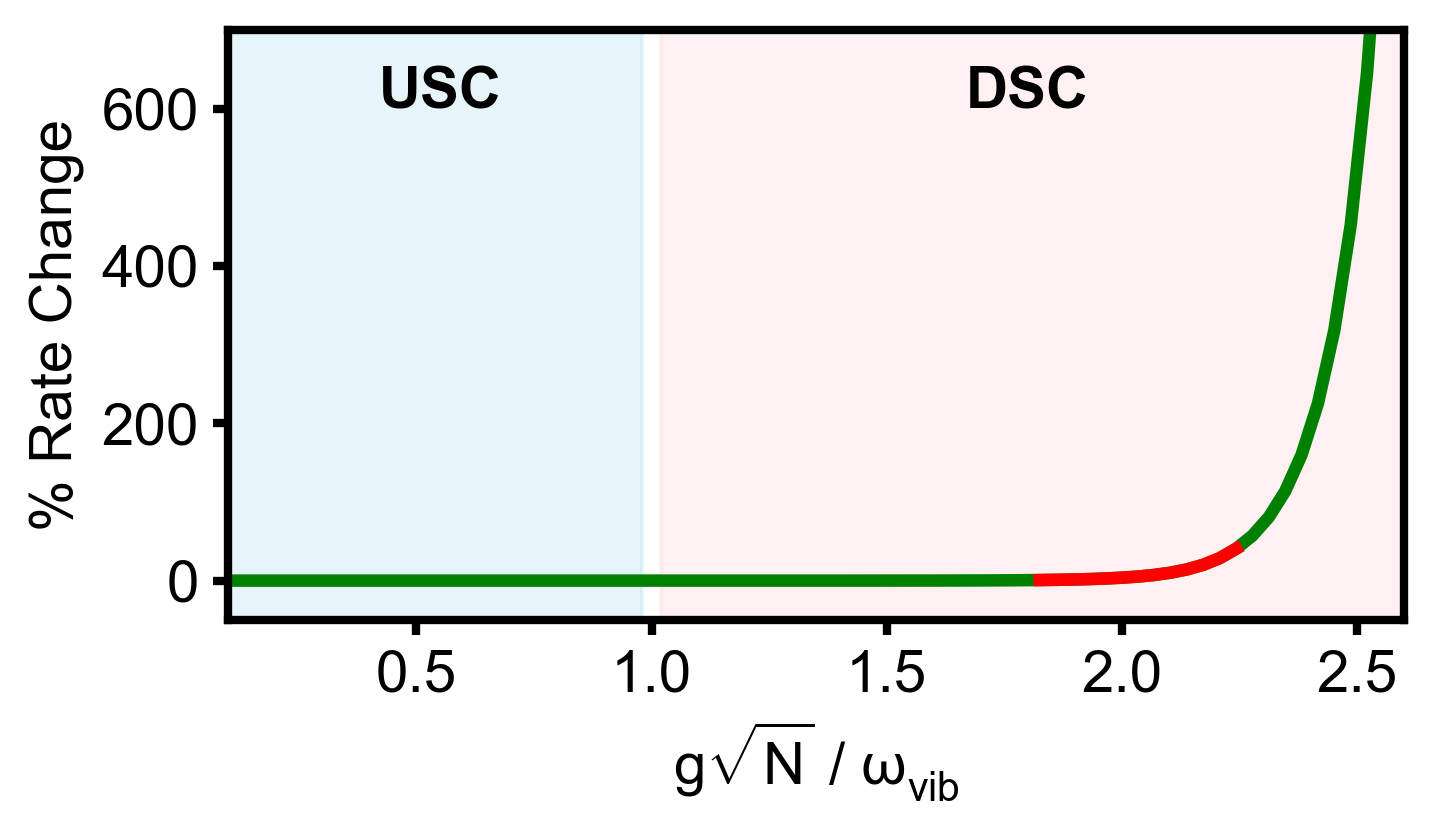}\caption{\label{fig:USC_condition}Rate changes due to VSC over a range of
collective light-matter coupling strengths $g\sqrt{N}/\omega_{\text{vib}}$.
Rate enhancements relative to the bare-molecule rate remain small
in the ultrastrong coupling regime (USC; characterised by $0.1\omega_{\text{vib}}\le g\sqrt{N}<\omega_{\text{vib}}$)
but become significant in the deep strong coupling regime (DSC; characterised
by $g\sqrt{N}\ge\omega_{\text{vib}}$). Note that the decay rates
under VSC were computed with $\bar{\bar{W}}_{\text{VSC}}^{\text{one}}$,
which applies the saddle point approximation to a different integral
form (see Supplementary Note 6). As such, these rates are accurate
only along the green curve (see Supplementary Note 8). All plots were
generated with the following parameters: detuning, $\Delta=0$; bare-molecule
Huang-Rhys factor, $S=0.1$; relative electronic energy gap, $\Delta E=15\hbar\omega_{\text{vib}}$;
number of molecules, $N=10^{10}$.}
\end{figure}

One way to achieve higher $\bar{\omega}_{\text{UP}}$ would be through
stronger light-matter couplings such as those in the ultrastrong regime
($0.1\omega_{\text{ph}}\le g\sqrt{N}<1$) and deep strong regime ($\omega_{\text{ph}}\le g\sqrt{N}$).
In these regimes, the RWA is no longer valid and corrections to the
polariton modes have to be made \citep{Rokaj2018,Mandal2020,CohenTannoudji1989,Rzazewski1975}.
The approach closely follows the work by Hopfield \citep{Hopfield1958}
and has been detailed in Supplementary Note 7; here, we summarise
the key results. For simplicity, we consider the zero detuning case
($\Delta=0$), such that $\omega_{\text{ph}}=\omega_{\text{vib}}$
and we may categorise coupling regimes with $\frac{g\sqrt{N}}{\omega_{\text{vib}}}$
as well. The corrected polariton modes have frequencies
\begin{align}
\omega_{\text{LP}} & =\omega_{\text{vib}}\br{\xi-\epsilon},\nonumber \\
\omega_{\text{UP}} & =\omega_{\text{vib}}\br{\xi+\epsilon},
\end{align}
with $\epsilon=g\sqrt{N}/\omega_{\text{vib}}$ and $\xi=\sqrt{1+\epsilon^{2}}$,
and Huang-Rhys factors
\begin{align}
S_{\text{LP}} & =\frac{S}{2N}\sigma_{\text{LP}},\nonumber \\
S_{\text{UP}} & =\frac{S}{2N}\sigma_{\text{UP}},
\end{align}
with $\sigma_{\text{LP}}=2K_{-}^{2}\sqbr{2\epsilon\br{\xi-\epsilon}-\br{\xi-\epsilon+1}}^{2}$
and $\sigma_{\text{UP}}=2K_{+}^{2}\sqbr{2\epsilon\br{\xi+\epsilon}+\br{\xi+\epsilon+1}}^{2}$,
where $\frac{1}{K_{\pm}^{2}}=2\sqbr{\pm\epsilon^{3}+\br{1\pm\epsilon}\br{1+\xi}^{2}+\epsilon^{2}\br{1+2\xi}}$.
Substituting these into Eq. (\ref{eq:VSC_condition-large_UP_freq-1}),
we obtain the new condition for polariton-mediated rate enhancement
\begin{align}
N & \lesssim\frac{1}{5}\frac{\Delta E}{\hbar\omega_{\text{vib}}}e^{\gamma\epsilon}e^{\gamma\br{\xi-1}}\sigma_{\text{UP}},\label{eq:USC_condition}
\end{align}
with the last two terms being corrections from the counter-rotating
and dipole self-energy terms. Note that $\xi-1$ and $\sigma_{\text{UP}}$
are positive, monotonically increasing functions with respect to $\epsilon$.
As such, the corrected UP mode will have higher frequencies and larger
effective Huang-Rhys factors (than the RWA case) that, ultimately,
make it easier to achieve rate enhancement. For instance, for a typical
system ($\Delta=0$, $S=0.1$, $\Delta E=15\hbar\omega_{\text{vib}}$)
with reasonable $N$ of $10^{10}$, Eq. (\ref{eq:USC_condition})
simplifies to $\frac{g\sqrt{N}}{\omega_{\text{vib}}}\gtrsim2.85$,
as opposed to $\frac{g\sqrt{N}}{\omega_{\text{vib}}}\gtrsim5.47$
from Eq. (\ref{eq:VSC_condition-large_UP_freq}) (i.e. under the RWA).
Regardless, decay through the UP is only significant in the deep strong
coupling regime. Numerical simulations suggest that over two-fold
rate enhancement may occur as early as $\frac{g\sqrt{N}}{\omega_{\text{vib}}}\gtrapprox2.5$
(Fig. \ref{fig:USC_condition}) and the computed decay rate is accurate
in this region. Systems of this regime, however, have not been realised
experimentally.

\subsection*{Effects of positive detunings on the non-radiative decay rate}

Perhaps a more practical way of achieving larger UP frequencies under
typical coupling regimes is to introduce some positive detuning $\Delta>0$.
If we assume $g\sqrt{N}$ to be small such that $g\sqrt{N}\ll\frac{\Delta}{2}$,
then the UP frequency is approximately $\omega_{\text{UP}}\approx\omega_{\text{vib}}+\Delta$
and can be increased with larger $\Delta$. This analysis is somewhat
naive because larger detunings make the UP mode more photon-like and
less matter-like, which in turn imposes a larger entropic penalty
on UP mode decay. Again, we hope for the frequency increase to dominate
over this greater entropy penalty. More specifically, we want
\begin{align}
N & \lesssim\frac{1}{5}\frac{\Delta E}{\hbar\omega_{\text{vib}}}e^{\gamma\br{\frac{\Delta}{\omega_{\text{vib}}}}}\br{2\sin^{2}\theta},
\end{align}
which is Eq. (\ref{eq:VSC_condition-large_UP_freq-1}) after substituting
$S_{\text{UP}}=\frac{S}{N}\sin^{2}\theta$ and $\bar{\omega}_{\text{UP}}\approx1+\frac{\Delta}{\omega_{\text{vib}}}$.
This translates to $\Delta\gtrsim8.7\omega_{\text{vib}}$ for a typical
system ($g\sqrt{N}=0.01\omega_{\text{vib}}$, $S=0.1$, $\Delta E=15\hbar\omega_{\text{vib}}$,
$N=10^{10}$).

Actually, this analysis is crude because the light-matter coupling
strength $g$ changes with $\Delta$. From cavity QED, the collective
light-matter coupling strength is \citep{Mandal2020}
\begin{align}
g\sqrt{N} & =\br{\sqrt{\frac{\omega_{\text{ph}}}{4\epsilon_{0}\mathcal{V}m\omega_{\text{vib}}}}\mu_{0}'}\sqrt{N},
\end{align}
where $\mathcal{V}$ is the effective quantisation volume of the cavity,
$m$ is the reduced mass of the single-molecule vibrational mode,
and $\mu_{0}'=\left.\frac{\partial\boldsymbol{\mu}\br x}{dx}\right|_{x=0}\cdot\boldsymbol{\epsilon}$
is the linear change in the dipole moment $\boldsymbol{\mu}$ along
the single-molecule vibrational mode, projected onto the polarisation
unit vector $\boldsymbol{\epsilon}$ (see Supplementary Note 7 for
more details). Also, for most experiments conducted in Fabry-Perot
cavities, $\omega_{\text{ph}}$ is modified by changing the distance
between the mirrors, so $\mathcal{V}\propto\frac{1}{\omega_{\text{ph}}}$.
Let us define the zero-detuning collective light-matter coupling strength
as $\tilde{g}\sqrt{N}=\br{\sqrt{\frac{\omega_{\text{vib}}}{4\epsilon_{0}\mathcal{V\omega_{\text{ph}}}m}}\mu_{0}'}\sqrt{N}$,
which is independent of $\omega_{\text{ph}}$ and is known to be $\approx0.01\omega_{\text{vib}}$.
Then, under off-resonant conditions, the collective light-matter coupling
strength becomes
\begin{align}
g\sqrt{N} & =\frac{\omega_{\text{ph}}}{\omega_{\text{vib}}}\tilde{g}\sqrt{N}=\br{\frac{\Delta}{\omega_{\text{vib}}}+1}\tilde{g}\sqrt{N},
\end{align}
which increases with detuning. Note that we need $g\sqrt{N}<0.1\omega_{\text{ph}}$
or $\tilde{g}\sqrt{N}<0.1\omega_{\text{vib}}$ for the RWA to remain
valid \citep{Rokaj2018,Mandal2020}. In other words, as we move towards
positive detunings, the RWA remains valid as long as this was the
case when the system was close to resonance.

\begin{figure}
\includegraphics[width=1\columnwidth]{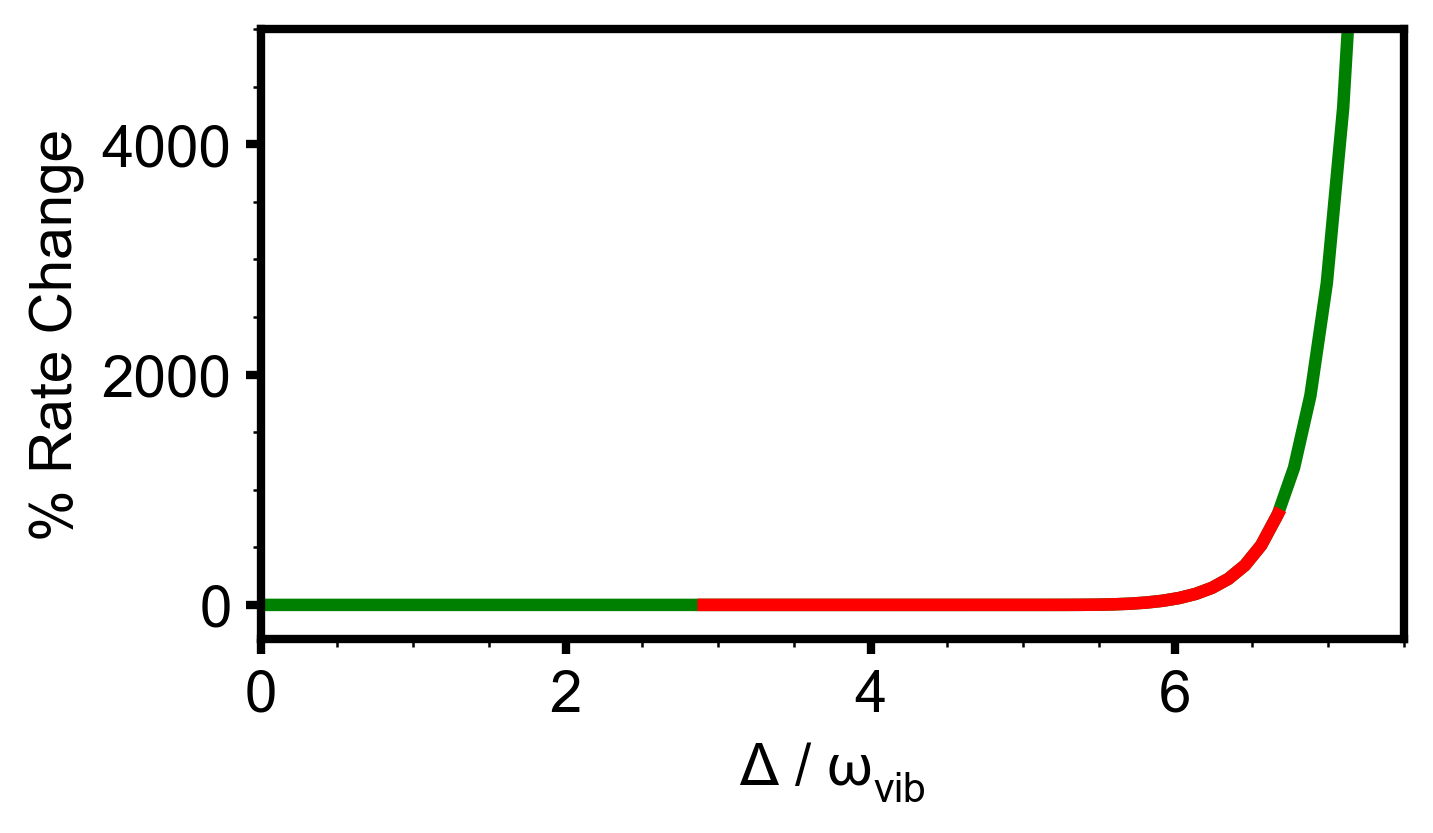}\caption{\label{fig:detuning_condition}Rate changes due to VSC over a range
of detunings $\Delta/\omega_{\text{vib}}$. Significant rate enhancements
are only observed at large positive detunings of $\Delta\gtrapprox7\omega_{\text{vib}}$.
Note that the decay rates under VSC were computed with $\bar{\bar{W}}_{\text{VSC}}^{\text{one}}$,
which applies the saddle point approximation to a different integral
form (see Supplementary Note 6). As such, these rates are accurate
only along the green curve (see Supplementary Note 8). All plots were
generated with the following parameters: zero-detuning collective
light-matter coupling, $\tilde{g}\sqrt{N}=0.01\omega_{\text{vib}}$;
bare-molecule Huang-Rhys factor, $S=0.1$; relative electronic energy
gap, $\Delta E=15\hbar\omega_{\text{vib}}$; number of molecules,
$N=10^{10}$.}
\end{figure}

Reformulating our numerical simulations with $\tilde{g}\sqrt{N}=0.01\omega_{\text{vib}}$
and, as before, $S=0.1$, $\Delta E=15\hbar\omega_{\text{vib}}$ and
$N=10^{10}$, we find more than 20-fold rate enhancements as early
as $\Delta\gtrapprox7\omega_{\text{vib}}$ or $\omega_{\text{ph}}\gtrapprox8\omega_{\text{vib}}$
(Fig. \ref{fig:detuning_condition}). The computed decay rate is accurate
in this region and agrees with our estimate of $\Delta\gtrsim8.7\omega_{\text{vib}}$.
Two comments are now in order:
\begin{itemize}
\item Firstly, at extremely large detunings of $\Delta\approx7\omega_{\text{vib}}$,
the UP mode is largely photonic. For instance, the Hopfield coefficient
of the UP, given by $\sin\theta$, is around $0.011$. This means
that rate enhancement is achieved by the creation of largely-photonic
quasiparticles. Since optical cavities are imperfect with a certain
degree of leakage, we wonder if these polaritons will result in photon
emission from the cavity. Clearly this non-radiative process is no
longer ``non-radiative'' under VSC because, even though the diabatic
coupling term $J_{\text{GE}}$ belongs to the matter component, these
processes may now emit quasiparticles that are primarily photons.
\item Secondly, there are plausibly other off-resonant effects that will
become crucial in this highly-detuned regime. For instance, a photon
energy of $8.0\hbar\omega_{\text{vib}}$ is more than half that of
the electronic energy gap $\Delta E=15\hbar\omega_{\text{vib}}$ and
may fall into the electronic strong coupling regime \citep{Ribeiro2018}.
Roughly-speaking, such systems have Hopfield coefficients of around
$0.11$ if we consider the lower exciton-polariton state in the single
excitation manifold (assuming a typical collective electronic light-matter
coupling strength $g_{\text{el}}\sqrt{N}$ of $0.05\frac{\Delta E}{\hbar}$).
This is not trivial when compared to the Hopfield coefficient of $0.011$
for the vibrational polariton mode; as such, our model is inaccurate
under these circumstances since both electronic and vibrational strong
couplings have to be considered.
\end{itemize}

\subsection*{Overcoming the entropic penalty with vibrational polariton condensates}

Polariton condensates provide a large number of initial excitations
in the polariton modes \citep{Proukakis2017} and have been proposed
as an avenue for improving chemical reactivities \citep{PannirSivajothi2022,Cortese2017,Phuc2022}.
In our case, a polariton condensate is expected to redistribute more
of the vibrational energy into the polariton modes so that decay through
these modes becomes more favourable. To illustrate this, we return
to the golden rule expression, but, this time, we assume a lower polariton
condensate with $N_{\text{LP}}$ excitations in the initial state
\begin{align}
 & W_{\text{cond}}\nonumber \\
 & =\frac{2\pi}{\hbar}\abs{J_{\text{GE}}}^{2}\sum_{\Delta n_{\text{LP}}=-N_{\text{LP}}}^{\infty}\sum_{n_{\text{UP}},n_{\text{D}}=0}^{\infty}\nonumber \\
 & \quad\delta\br{\Delta E-\Delta n_{\text{LP}}\hbar\omega_{\text{LP}}-n_{\text{UP}}\hbar\omega_{\text{UP}}-n_{\text{D}}\hbar\omega_{\text{vib}}}\nonumber \\
 & \quad\times\abs{\braket{n_{\text{LP}}^{\br{\text{LP}'}},n_{\text{UP}}^{\br{\text{UP}'}},n_{\text{D}}^{\br{\text{D}'}}\vert N_{\text{LP}}^{\br{\text{LP}}},0^{\br{\text{UP}}},0^{\br{\text{D}}}}}^{2},
\end{align}
where $W_{\text{cond}}$ is the non-radiative decay rate under VSC
with a LP condensate and $\Delta n_{\text{LP}}=n_{\text{LP}}-N_{\text{LP}}$
is an integer running from $-N_{\text{LP}}$ to $\infty$ and represents
the change in number of LP quanta after non-radiative decay. By writing
the Dirac delta function in its Fourier form, bringing only the sum
over final dark states into the integral and performing many saddle
point approximations, we obtain
\begin{align}
W_{\text{cond}} & \approx\frac{\abs{J_{\text{GE}}}^{2}}{\hbar}e^{-S}\nonumber \\
 & \quad\times\sum_{\Delta n_{\text{LP}}=-N_{\text{LP}}}^{\infty}\sum_{n_{\text{UP}}=0}^{\infty}w_{\text{cond}}\br{\Delta n_{\text{LP}},n_{\text{UP}}},\label{eq:W_cond}
\end{align}
with
\begin{align}
 & w_{\text{cond}}\br{\Delta n_{\text{LP}},n_{\text{UP}}}\nonumber \\
 & =\abs{e^{\frac{1}{2}S_{\text{LP}}}\braket{n_{\text{LP}}^{\br{\text{LP}'}}\vert N_{\text{LP}}^{\br{\text{LP}}}}}^{2}\frac{\br{S_{\text{UP}}}^{n_{\text{UP}}}}{n_{\text{UP}}!}\nonumber \\
 & \quad\times\sqrt{\frac{2\pi}{\hbar\omega_{\text{vib}}\Delta\tilde{E}_{\text{cond}}}}\nonumber \\
 & \quad\times\exp\sqbr{-\br{\ln\frac{\Delta\tilde{E}_{\text{cond}}}{S_{\text{D}}\hbar\omega_{\text{vib}}}-1}\frac{\Delta\tilde{E}_{\text{cond}}}{\hbar\omega_{\text{vib}}}},
\end{align}
where $\Delta\tilde{E}_{\text{cond}}=\Delta E-\Delta n_{\text{LP}}\hbar\omega_{\text{LP}}-n_{\text{UP}}\hbar\omega_{\text{UP}}$
is the new effective energy gap for dark mode decay. Again, the approximation
is valid in the limit of $\frac{\Delta\tilde{E}_{\text{cond}}}{\hbar\omega_{\text{vib}}}\gg1$.
This result has the same physical interpretation as $W_{\text{VSC}}^{\text{many}}$.

\begin{figure*}
\includegraphics[width=2\columnwidth]{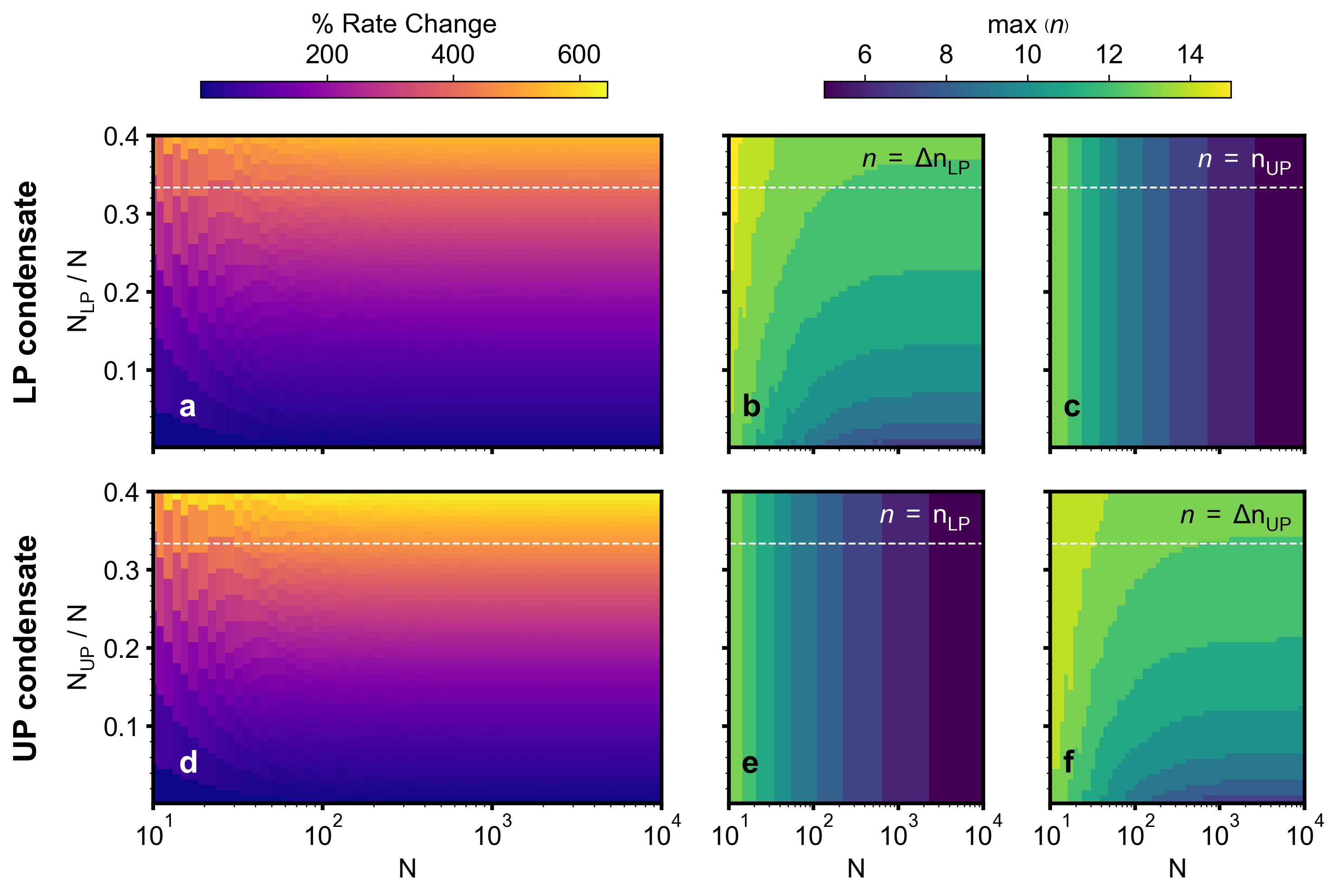}\caption{\label{fig:condensate_condition}Effect of polariton condensates on
non-radiative decay rate under VSC. (a) Rate changes due to VSC over
ranges of number of molecules $N$ and LP condensate ratios $N_{\text{LP}}/N$.
Larger values of $N_{\text{LP}}/N$ give higher rate changes that
are no longer dependent on $N$ for sufficiently large $N$. (b,c)
Contributions to the decay rate by the LP and UP modes. The value
$\max\protect\br n$ represents the number of $n$ terms ($n=\Delta n_{\text{LP}},n_{\text{UP}}$)
in Eq. (\ref{eq:W_cond}) that needed to be summed before the subsequent
term falls below $10^{-15}$. Changes in rate due to $N_{\text{LP}}/N$
may be attributed to changes in $\max\protect\br{\Delta n_{\text{LP}}}$,
both of which remain constant over $N$ for sufficiently large $N$.
Meanwhile, $\max\protect\br{n_{\text{UP}}}$ remains constant with
increasing $N_{\text{LP}}/N$. All these indicate that a LP condensate
speeds up non-radiative decay through the LP mode and not the UP mode.
(d,e,f) Same plots as (a,b,c) but with an UP condensate. Both types
of condensates follow similar trends, with the UP condensate giving
higher decay rates due to the higher frequency associated with the
UP mode. All plots were generated with the following parameters: detuning,
$\Delta=0$; collective light-matter coupling, $g\sqrt{N}=0.01\omega_{\text{vib}}$;
bare-molecule Huang-Rhys factor, $S=0.1$; relative electronic energy
gap, $\Delta E=15\hbar\omega_{\text{vib}}$. Dotted lines represent
$\frac{N_{P}}{N}=5\frac{\hbar\omega_{\text{vib}}}{\Delta E}\approx0.33$,
with $P=\text{LP},\text{UP}$.}
\end{figure*}

By estimating the excited-state Franck-Condon overlap \citep{Palma1983}
(see Supplementary Note 9), the $\br{\Delta n_{\text{LP}},n_{\text{UP}}}=\br{0,0}$
term in Eq. (\ref{eq:W_cond}) reduces to the single molecule decay
rate in the large $N$ limit, signifying that terms of $\Delta n_{\text{LP}},n_{\text{UP}}\neq0$
represent the polariton contributions to the decay rate. The UP contribution
may be represented by $w_{\text{cond}}\br{0,1}\approx S_{\text{UP}}e^{\gamma}w_{\text{cond}}\br{0,0}$,
where $\gamma=\ln\frac{\Delta E}{S_{\text{D}}\hbar\omega_{\text{vib}}}-1$.
This equation takes the same form as Eq. (\ref{eq:VSC_condition-1})
and may be reduced to Eq. (\ref{eq:VSC_condition}): the small $N$
condition for polariton-mediated rate enhancement. What is interesting,
however, is the decay through the LP mode under the influence of a
LP condensate. We compute
\begin{align}
w_{\text{cond}}\br{1,0} & \approx S_{\text{LP}}\br{N_{\text{LP}}+1}e^{\gamma}w_{\text{cond}}\br{0,0},
\end{align}
and find that the condensate effectively increases the LP Huang-Rhys
factor from $S_{\text{LP}}$ to $S_{\text{LP}}\br{N_{\text{LP}}+1}$.
Finally, decay through the LP mode becomes significant when $w_{\text{cond}}\br{1,0}>w_{\text{cond}}\br{0,0}$,
or
\begin{align}
\frac{N_{\text{LP}}}{N} & \gtrsim5\frac{\hbar\omega_{\text{vib}}}{\Delta E},\label{eq:condensate_condition}
\end{align}
where we have noted that, near resonance, $S_{\text{LP}}\simeq\frac{S}{2N}$.
Physically, the LP condensate supplies energy to the LP mode, giving
it more entropic ``weight'' during non-radiative decay. This effect
is seen by an increase in the Huang-Rhys factor from $\frac{S}{2}\br{\frac{1}{N}}$
to $\frac{S}{2}\br{\frac{N_{\text{LP}}}{N}}$, thereby providing a
smaller ``effective'' $N$ needed in the small $N$ condition of
Eq. (\ref{eq:VSC_condition}). Running numerical simulations with
the same conditions of $\Delta=0$, $g\sqrt{N}=0.01\omega_{\text{vib}}$,
$S=0.1$ and a specific $\Delta E=15\hbar\omega_{\text{vib}}$, we
see in Fig. \ref{fig:condensate_condition}a that, with $\frac{N_{\text{LP}}}{N}\gtrsim5\frac{\hbar\omega_{\text{vib}}}{\Delta E}\approx0.33$,
appreciable rate enhancements may be observed even at large $N$ of
$10^{4}$. We may also verify in Fig. \ref{fig:condensate_condition}b
and \ref{fig:condensate_condition}c that the LP is indeed responsible
for this effect. Thus, rate enhancement is now possible even for more
realistic systems with smaller energy gaps; for instance, with $\Delta E=7\hbar\omega_{\text{vib}}\approx21000\mathrm{\,cm^{-1}}$.
we need a condensate ratio of $\frac{N_{\text{LP}}}{N}\gtrsim0.7$.
Note that the condensate ratios considered by this work were estimated
based on the fraction of excitations that can be achieved in a nonlinear
optical spectroscopy experiment, which has been estimated by Ribeiro
and co-workers to be $\mathcal{O}\br{10^{-1}}$ \citep{Ribeiro2018-pump}.
As an aside, we have chosen $\Delta n_{\text{LP}}=+1$ and not $-1$
to obtain Eq. (\ref{eq:condensate_condition}) because the latter
would increase the effective energy gap $\Delta\tilde{E}_{\text{cond}}$
for dark mode decay and reduce the overall decay rate.

Similar arguments may be made for an upper polariton condensate with
$N_{\text{UP}}$ excitations in the initial state. By mapping $\text{LP}\rightarrow\text{UP}$,
the condensate's effects can be felt when $\frac{N_{\text{UP}}}{N}\gtrsim5\frac{\hbar\omega_{\text{vib}}}{\Delta E}$,
which is verified by performing numerical simulations using the same
set of parameters as before (Fig. \ref{fig:condensate_condition}d-f).
Of course, the UP mode has a higher frequency than the LP mode, so
the UP condensate gives faster decay rates than the LP condensate.
Other than that, similar trends are observed for both types of condensates.

\section*{Conclusions}

Collective VSC is an ensemble effect, that is, we need a macroscopic
number of molecules $N$ to couple to the cavity mode for the light-matter
coupling to be appreciable. Where light-matter interactions are concerned,
one reacting molecule is only a needle in a haystack: only $\frac{1}{N}$
of the vibrational mode couples to the photon mode (through the polaritons)
while the remaining $\frac{N-1}{N}$ of it is distributed into the
dark modes that are chemically identical to the bare molecule (at
least in the absence of disorder and non-equilibrium effects \citep{Du2022});
this is the essence of the polariton ``large $N$ problem'' \citep{MartinezMartinez2019}.
As such, we do not expect VSC to modify the non-radiative decay rate
unless the polariton modes possess some unique advantage over the
dark modes. In this paper, we show how the higher UP frequency (relative
to the dark mode) is this advantage, yet it cannot be appreciably
exploited under most circumstances for VSC. Instead, we require conditions
that are either not yet experimentally achievable (deep strong light-matter
coupling) or associated with theoretical complications (highly off-resonant
cavities). Vibrational polariton condensates, on the other hand, may
provide an elegant solution to this entropy problem. By starting from
a macroscopic occupation of a polariton mode, more of the vibrational
energy may be redistributed through the polariton modes. Care must
be taken in this regime though, since, under such extreme conditions,
one may experience other effects, such as chemical reactions and other
transfer processes, that have not been considered in this model. Regardless
of the mechanism involved, once polaritons start affecting non-radiative
decay rates, this process is no longer fully ``non-radiative'' since
the quasiparticles produced have some light component in them. Moving
forward, it would be interesting to investigate whether electronically
excited molecules under VSC conditions may emit infrared photons out
of the cavity via previously non-radiative pathways. This could potentially
lead to new molecular optoelectronic applications.

\section*{Acknowledgements}

Y.R.P. was supported by the UCSD Graduate Student Growth and Excellent
Initiative Model. S.P.-S. and J.Y-Z. were supported by the US Department
of Energy, Office of Science, Basic Energy Sciences, CPIMS Program
under Early Career Research Program Award DE-SC0019188. We also thank
Jorge A. Campos-Gonzalez-Angulo for helpful discussions.


\begin{thebibliography}{63}%
\makeatletter
\providecommand \@ifxundefined [1]{%
 \@ifx{#1\undefined}
}%
\providecommand \@ifnum [1]{%
 \ifnum #1\expandafter \@firstoftwo
 \else \expandafter \@secondoftwo
 \fi
}%
\providecommand \@ifx [1]{%
 \ifx #1\expandafter \@firstoftwo
 \else \expandafter \@secondoftwo
 \fi
}%
\providecommand \natexlab [1]{#1}%
\providecommand \enquote  [1]{``#1''}%
\providecommand \bibnamefont  [1]{#1}%
\providecommand \bibfnamefont [1]{#1}%
\providecommand \citenamefont [1]{#1}%
\providecommand \href@noop [0]{\@secondoftwo}%
\providecommand \href [0]{\begingroup \@sanitize@url \@href}%
\providecommand \@href[1]{\@@startlink{#1}\@@href}%
\providecommand \@@href[1]{\endgroup#1\@@endlink}%
\providecommand \@sanitize@url [0]{\catcode `\\12\catcode `\$12\catcode
  `\&12\catcode `\#12\catcode `\^12\catcode `\_12\catcode `\%12\relax}%
\providecommand \@@startlink[1]{}%
\providecommand \@@endlink[0]{}%
\providecommand \url  [0]{\begingroup\@sanitize@url \@url }%
\providecommand \@url [1]{\endgroup\@href {#1}{\urlprefix }}%
\providecommand \urlprefix  [0]{URL }%
\providecommand \Eprint [0]{\href }%
\providecommand \doibase [0]{https://doi.org/}%
\providecommand \selectlanguage [0]{\@gobble}%
\providecommand \bibinfo  [0]{\@secondoftwo}%
\providecommand \bibfield  [0]{\@secondoftwo}%
\providecommand \translation [1]{[#1]}%
\providecommand \BibitemOpen [0]{}%
\providecommand \bibitemStop [0]{}%
\providecommand \bibitemNoStop [0]{.\EOS\space}%
\providecommand \EOS [0]{\spacefactor3000\relax}%
\providecommand \BibitemShut  [1]{\csname bibitem#1\endcsname}%
\let\auto@bib@innerbib\@empty
%</preamble>
\bibitem [{\citenamefont {Lidzey}\ \emph {et~al.}(1998)\citenamefont {Lidzey},
  \citenamefont {Bradley}, \citenamefont {Skolnick}, \citenamefont {Virgili},
  \citenamefont {Walker},\ and\ \citenamefont {Whittaker}}]{Lidzey1998}%
  \BibitemOpen
  \bibfield  {author} {\bibinfo {author} {\bibfnamefont {D.~G.}\ \bibnamefont
  {Lidzey}}, \bibinfo {author} {\bibfnamefont {D.~D.~C.}\ \bibnamefont
  {Bradley}}, \bibinfo {author} {\bibfnamefont {M.~S.}\ \bibnamefont
  {Skolnick}}, \bibinfo {author} {\bibfnamefont {T.}~\bibnamefont {Virgili}},
  \bibinfo {author} {\bibfnamefont {S.}~\bibnamefont {Walker}},\ and\ \bibinfo
  {author} {\bibfnamefont {D.~M.}\ \bibnamefont {Whittaker}},\ }\href
  {https://doi.org/10.1038/25692} {\bibfield  {journal} {\bibinfo  {journal}
  {Nature}\ }\textbf {\bibinfo {volume} {395}},\ \bibinfo {pages} {53}
  (\bibinfo {year} {1998})}\BibitemShut {NoStop}%
\bibitem [{\citenamefont {Long}\ and\ \citenamefont
  {Simpkins}(2015)}]{Long2015}%
  \BibitemOpen
  \bibfield  {author} {\bibinfo {author} {\bibfnamefont {J.~P.}\ \bibnamefont
  {Long}}\ and\ \bibinfo {author} {\bibfnamefont {B.~S.}\ \bibnamefont
  {Simpkins}},\ }\href {https://doi.org/10.1021/ph5003347} {\bibfield
  {journal} {\bibinfo  {journal} {ACS Photonics}\ }\textbf {\bibinfo {volume}
  {2}},\ \bibinfo {pages} {130} (\bibinfo {year} {2015})}\BibitemShut {NoStop}%
\bibitem [{\citenamefont {Shalabney}\ \emph {et~al.}(2015)\citenamefont
  {Shalabney}, \citenamefont {George}, \citenamefont {Hutchison}, \citenamefont
  {Pupillo}, \citenamefont {Genet},\ and\ \citenamefont
  {Ebbesen}}]{Shalabney2015}%
  \BibitemOpen
  \bibfield  {author} {\bibinfo {author} {\bibfnamefont {A.}~\bibnamefont
  {Shalabney}}, \bibinfo {author} {\bibfnamefont {J.}~\bibnamefont {George}},
  \bibinfo {author} {\bibfnamefont {J.}~\bibnamefont {Hutchison}}, \bibinfo
  {author} {\bibfnamefont {G.}~\bibnamefont {Pupillo}}, \bibinfo {author}
  {\bibfnamefont {C.}~\bibnamefont {Genet}},\ and\ \bibinfo {author}
  {\bibfnamefont {T.~W.}\ \bibnamefont {Ebbesen}},\ }\href
  {https://doi.org/10.1038/ncomms6981} {\bibfield  {journal} {\bibinfo
  {journal} {Nature Communications}\ }\textbf {\bibinfo {volume} {6}},\
  \bibinfo {pages} {5981} (\bibinfo {year} {2015})}\BibitemShut {NoStop}%
\bibitem [{\citenamefont {Galego}\ \emph {et~al.}(2015)\citenamefont {Galego},
  \citenamefont {Garcia-Vidal},\ and\ \citenamefont {Feist}}]{Galego2015}%
  \BibitemOpen
  \bibfield  {author} {\bibinfo {author} {\bibfnamefont {J.}~\bibnamefont
  {Galego}}, \bibinfo {author} {\bibfnamefont {F.~J.}\ \bibnamefont
  {Garcia-Vidal}},\ and\ \bibinfo {author} {\bibfnamefont {J.}~\bibnamefont
  {Feist}},\ }\href {https://doi.org/10.1103/PhysRevX.5.041022} {\bibfield
  {journal} {\bibinfo  {journal} {Physical Review X}\ }\textbf {\bibinfo
  {volume} {5}},\ \bibinfo {pages} {041022} (\bibinfo {year}
  {2015})}\BibitemShut {NoStop}%
\bibitem [{\citenamefont {Feist}\ \emph {et~al.}(2018)\citenamefont {Feist},
  \citenamefont {Galego},\ and\ \citenamefont {Garcia-Vidal}}]{Feist2018}%
  \BibitemOpen
  \bibfield  {author} {\bibinfo {author} {\bibfnamefont {J.}~\bibnamefont
  {Feist}}, \bibinfo {author} {\bibfnamefont {J.}~\bibnamefont {Galego}},\ and\
  \bibinfo {author} {\bibfnamefont {F.~J.}\ \bibnamefont {Garcia-Vidal}},\
  }\href {https://doi.org/10.1021/acsphotonics.7b00680} {\bibfield  {journal}
  {\bibinfo  {journal} {ACS Photonics}\ }\textbf {\bibinfo {volume} {5}},\
  \bibinfo {pages} {205} (\bibinfo {year} {2018})}\BibitemShut {NoStop}%
\bibitem [{\citenamefont {Ebbesen}(2016)}]{Ebbesen2016}%
  \BibitemOpen
  \bibfield  {author} {\bibinfo {author} {\bibfnamefont {T.~W.}\ \bibnamefont
  {Ebbesen}},\ }\href {https://doi.org/10.1021/acs.accounts.6b00295} {\bibfield
   {journal} {\bibinfo  {journal} {Accounts of Chemical Research}\ }\textbf
  {\bibinfo {volume} {49}},\ \bibinfo {pages} {2403} (\bibinfo {year}
  {2016})}\BibitemShut {NoStop}%
\bibitem [{\citenamefont {Thomas}\ \emph {et~al.}(2016)\citenamefont {Thomas},
  \citenamefont {George}, \citenamefont {Shalabney}, \citenamefont {Dryzhakov},
  \citenamefont {Varma}, \citenamefont {Moran}, \citenamefont {Chervy},
  \citenamefont {Zhong}, \citenamefont {Devaux}, \citenamefont {Genet},
  \citenamefont {Hutchison},\ and\ \citenamefont {Ebbesen}}]{Thomas2016}%
  \BibitemOpen
  \bibfield  {author} {\bibinfo {author} {\bibfnamefont {A.}~\bibnamefont
  {Thomas}}, \bibinfo {author} {\bibfnamefont {J.}~\bibnamefont {George}},
  \bibinfo {author} {\bibfnamefont {A.}~\bibnamefont {Shalabney}}, \bibinfo
  {author} {\bibfnamefont {M.}~\bibnamefont {Dryzhakov}}, \bibinfo {author}
  {\bibfnamefont {S.~J.}\ \bibnamefont {Varma}}, \bibinfo {author}
  {\bibfnamefont {J.}~\bibnamefont {Moran}}, \bibinfo {author} {\bibfnamefont
  {T.}~\bibnamefont {Chervy}}, \bibinfo {author} {\bibfnamefont
  {X.}~\bibnamefont {Zhong}}, \bibinfo {author} {\bibfnamefont
  {E.}~\bibnamefont {Devaux}}, \bibinfo {author} {\bibfnamefont
  {C.}~\bibnamefont {Genet}}, \bibinfo {author} {\bibfnamefont {J.~A.}\
  \bibnamefont {Hutchison}},\ and\ \bibinfo {author} {\bibfnamefont {T.~W.}\
  \bibnamefont {Ebbesen}},\ }\href {https://doi.org/10.1002/anie.201605504}
  {\bibfield  {journal} {\bibinfo  {journal} {Angewandte Chemie International
  Edition}\ }\textbf {\bibinfo {volume} {55}},\ \bibinfo {pages} {11462}
  (\bibinfo {year} {2016})}\BibitemShut {NoStop}%
\bibitem [{\citenamefont {Pang}\ \emph {et~al.}(2020)\citenamefont {Pang},
  \citenamefont {Thomas}, \citenamefont {Nagarajan}, \citenamefont {Vergauwe},
  \citenamefont {Joseph}, \citenamefont {Patrahau}, \citenamefont {Wang},
  \citenamefont {Genet},\ and\ \citenamefont {Ebbesen}}]{Pang2020}%
  \BibitemOpen
  \bibfield  {author} {\bibinfo {author} {\bibfnamefont {Y.}~\bibnamefont
  {Pang}}, \bibinfo {author} {\bibfnamefont {A.}~\bibnamefont {Thomas}},
  \bibinfo {author} {\bibfnamefont {K.}~\bibnamefont {Nagarajan}}, \bibinfo
  {author} {\bibfnamefont {R.~M.~A.}\ \bibnamefont {Vergauwe}}, \bibinfo
  {author} {\bibfnamefont {K.}~\bibnamefont {Joseph}}, \bibinfo {author}
  {\bibfnamefont {B.}~\bibnamefont {Patrahau}}, \bibinfo {author}
  {\bibfnamefont {K.}~\bibnamefont {Wang}}, \bibinfo {author} {\bibfnamefont
  {C.}~\bibnamefont {Genet}},\ and\ \bibinfo {author} {\bibfnamefont {T.~W.}\
  \bibnamefont {Ebbesen}},\ }\href {https://doi.org/10.1002/anie.202002527}
  {\bibfield  {journal} {\bibinfo  {journal} {Angewandte Chemie International
  Edition}\ }\textbf {\bibinfo {volume} {59}},\ \bibinfo {pages} {10436}
  (\bibinfo {year} {2020})}\BibitemShut {NoStop}%
\bibitem [{\citenamefont {Garcia-Vidal}\ \emph {et~al.}(2021)\citenamefont
  {Garcia-Vidal}, \citenamefont {Ciuti},\ and\ \citenamefont
  {Ebbesen}}]{GarciaVidal2021}%
  \BibitemOpen
  \bibfield  {author} {\bibinfo {author} {\bibfnamefont {F.~J.}\ \bibnamefont
  {Garcia-Vidal}}, \bibinfo {author} {\bibfnamefont {C.}~\bibnamefont
  {Ciuti}},\ and\ \bibinfo {author} {\bibfnamefont {T.~W.}\ \bibnamefont
  {Ebbesen}},\ }\href {https://doi.org/10.1126/science.abd0336} {\bibfield
  {journal} {\bibinfo  {journal} {Science}\ }\textbf {\bibinfo {volume}
  {373}},\ \bibinfo {pages} {eabd0336} (\bibinfo {year} {2021})}\BibitemShut
  {NoStop}%
\bibitem [{\citenamefont {Thomas}\ \emph {et~al.}(2019)\citenamefont {Thomas},
  \citenamefont {Lethuillier-Karl}, \citenamefont {Nagarajan}, \citenamefont
  {Vergauwe}, \citenamefont {George}, \citenamefont {Chervy}, \citenamefont
  {Shalabney}, \citenamefont {Devaux}, \citenamefont {Genet}, \citenamefont
  {Moran},\ and\ \citenamefont {Ebbesen}}]{Thomas2019}%
  \BibitemOpen
  \bibfield  {author} {\bibinfo {author} {\bibfnamefont {A.}~\bibnamefont
  {Thomas}}, \bibinfo {author} {\bibfnamefont {L.}~\bibnamefont
  {Lethuillier-Karl}}, \bibinfo {author} {\bibfnamefont {K.}~\bibnamefont
  {Nagarajan}}, \bibinfo {author} {\bibfnamefont {R.~M.~A.}\ \bibnamefont
  {Vergauwe}}, \bibinfo {author} {\bibfnamefont {J.}~\bibnamefont {George}},
  \bibinfo {author} {\bibfnamefont {T.}~\bibnamefont {Chervy}}, \bibinfo
  {author} {\bibfnamefont {A.}~\bibnamefont {Shalabney}}, \bibinfo {author}
  {\bibfnamefont {E.}~\bibnamefont {Devaux}}, \bibinfo {author} {\bibfnamefont
  {C.}~\bibnamefont {Genet}}, \bibinfo {author} {\bibfnamefont
  {J.}~\bibnamefont {Moran}},\ and\ \bibinfo {author} {\bibfnamefont {T.~W.}\
  \bibnamefont {Ebbesen}},\ }\href {https://doi.org/10.1126/science.aau7742}
  {\bibfield  {journal} {\bibinfo  {journal} {Science}\ }\textbf {\bibinfo
  {volume} {363}},\ \bibinfo {pages} {615} (\bibinfo {year}
  {2019})}\BibitemShut {NoStop}%
\bibitem [{\citenamefont {Hirai}\ \emph {et~al.}(2020)\citenamefont {Hirai},
  \citenamefont {Takeda}, \citenamefont {Hutchison},\ and\ \citenamefont
  {Uji-i}}]{Hirai2020}%
  \BibitemOpen
  \bibfield  {author} {\bibinfo {author} {\bibfnamefont {K.}~\bibnamefont
  {Hirai}}, \bibinfo {author} {\bibfnamefont {R.}~\bibnamefont {Takeda}},
  \bibinfo {author} {\bibfnamefont {J.~A.}\ \bibnamefont {Hutchison}},\ and\
  \bibinfo {author} {\bibfnamefont {H.}~\bibnamefont {Uji-i}},\ }\href
  {https://doi.org/10.1002/anie.201915632} {\bibfield  {journal} {\bibinfo
  {journal} {Angewandte Chemie International Edition}\ }\textbf {\bibinfo
  {volume} {59}},\ \bibinfo {pages} {5332} (\bibinfo {year}
  {2020})}\BibitemShut {NoStop}%
\bibitem [{\citenamefont {Campos-Gonzalez-Angulo}\ and\ \citenamefont
  {Yuen-Zhou}(2020)}]{CamposGonzalezAngulo2020}%
  \BibitemOpen
  \bibfield  {author} {\bibinfo {author} {\bibfnamefont {J.~A.}\ \bibnamefont
  {Campos-Gonzalez-Angulo}}\ and\ \bibinfo {author} {\bibfnamefont
  {J.}~\bibnamefont {Yuen-Zhou}},\ }\href {https://doi.org/10.1063/5.0007547}
  {\bibfield  {journal} {\bibinfo  {journal} {The Journal of Chemical Physics}\
  }\textbf {\bibinfo {volume} {152}},\ \bibinfo {pages} {161101} (\bibinfo
  {year} {2020})}\BibitemShut {NoStop}%
\bibitem [{\citenamefont {Botzung}\ \emph {et~al.}(2020)\citenamefont
  {Botzung}, \citenamefont {Hagenmüller}, \citenamefont {Schütz},
  \citenamefont {Dubail}, \citenamefont {Pupillo},\ and\ \citenamefont
  {Schachenmayer}}]{Botzung2020}%
  \BibitemOpen
  \bibfield  {author} {\bibinfo {author} {\bibfnamefont {T.}~\bibnamefont
  {Botzung}}, \bibinfo {author} {\bibfnamefont {D.}~\bibnamefont
  {Hagenmüller}}, \bibinfo {author} {\bibfnamefont {S.}~\bibnamefont
  {Schütz}}, \bibinfo {author} {\bibfnamefont {J.}~\bibnamefont {Dubail}},
  \bibinfo {author} {\bibfnamefont {G.}~\bibnamefont {Pupillo}},\ and\ \bibinfo
  {author} {\bibfnamefont {J.}~\bibnamefont {Schachenmayer}},\ }\href
  {https://doi.org/10.1103/PhysRevB.102.144202} {\bibfield  {journal} {\bibinfo
   {journal} {Physical Review B}\ }\textbf {\bibinfo {volume} {102}},\ \bibinfo
  {pages} {144202} (\bibinfo {year} {2020})}\BibitemShut {NoStop}%
\bibitem [{\citenamefont {Scholes}(2020)}]{Scholes2020}%
  \BibitemOpen
  \bibfield  {author} {\bibinfo {author} {\bibfnamefont {G.~D.}\ \bibnamefont
  {Scholes}},\ }\href {https://doi.org/10.1098/rspa.2020.0278} {\bibfield
  {journal} {\bibinfo  {journal} {Proceedings of the Royal Society A:
  Mathematical, Physical and Engineering Sciences}\ }\textbf {\bibinfo {volume}
  {476}},\ \bibinfo {pages} {20200278} (\bibinfo {year} {2020})}\BibitemShut
  {NoStop}%
\bibitem [{\citenamefont {Du}\ and\ \citenamefont {Yuen-Zhou}(2022)}]{Du2022}%
  \BibitemOpen
  \bibfield  {author} {\bibinfo {author} {\bibfnamefont {M.}~\bibnamefont
  {Du}}\ and\ \bibinfo {author} {\bibfnamefont {J.}~\bibnamefont {Yuen-Zhou}},\
  }\href {https://doi.org/10.1103/PhysRevLett.128.096001} {\bibfield  {journal}
  {\bibinfo  {journal} {Physical Review Letters}\ }\textbf {\bibinfo {volume}
  {128}},\ \bibinfo {pages} {096001} (\bibinfo {year} {2022})}\BibitemShut
  {NoStop}%
\bibitem [{\citenamefont {Campos-Gonzalez-Angulo}\ \emph
  {et~al.}(2019)\citenamefont {Campos-Gonzalez-Angulo}, \citenamefont
  {Ribeiro},\ and\ \citenamefont {Yuen-Zhou}}]{CamposGonzalezAngulo2019}%
  \BibitemOpen
  \bibfield  {author} {\bibinfo {author} {\bibfnamefont {J.~A.}\ \bibnamefont
  {Campos-Gonzalez-Angulo}}, \bibinfo {author} {\bibfnamefont {R.~F.}\
  \bibnamefont {Ribeiro}},\ and\ \bibinfo {author} {\bibfnamefont
  {J.}~\bibnamefont {Yuen-Zhou}},\ }\href
  {https://doi.org/10.1038/s41467-019-12636-1} {\bibfield  {journal} {\bibinfo
  {journal} {Nature Communications}\ }\textbf {\bibinfo {volume} {10}},\
  \bibinfo {pages} {4685} (\bibinfo {year} {2019})}\BibitemShut {NoStop}%
\bibitem [{\citenamefont {Pannir-Sivajothi}\ \emph {et~al.}(2022)\citenamefont
  {Pannir-Sivajothi}, \citenamefont {Campos-Gonzalez-Angulo}, \citenamefont
  {Martínez-Martínez}, \citenamefont {Sinha},\ and\ \citenamefont
  {Yuen-Zhou}}]{PannirSivajothi2022}%
  \BibitemOpen
  \bibfield  {author} {\bibinfo {author} {\bibfnamefont {S.}~\bibnamefont
  {Pannir-Sivajothi}}, \bibinfo {author} {\bibfnamefont {J.~A.}\ \bibnamefont
  {Campos-Gonzalez-Angulo}}, \bibinfo {author} {\bibfnamefont {L.~A.}\
  \bibnamefont {Martínez-Martínez}}, \bibinfo {author} {\bibfnamefont
  {S.}~\bibnamefont {Sinha}},\ and\ \bibinfo {author} {\bibfnamefont
  {J.}~\bibnamefont {Yuen-Zhou}},\ }\href
  {https://doi.org/10.1038/s41467-022-29290-9} {\bibfield  {journal} {\bibinfo
  {journal} {Nature Communications}\ }\textbf {\bibinfo {volume} {13}},\
  \bibinfo {pages} {1645} (\bibinfo {year} {2022})}\BibitemShut {NoStop}%
\bibitem [{\citenamefont {Cortese}\ \emph {et~al.}(2017)\citenamefont
  {Cortese}, \citenamefont {Lagoudakis},\ and\ \citenamefont
  {De~Liberato}}]{Cortese2017}%
  \BibitemOpen
  \bibfield  {author} {\bibinfo {author} {\bibfnamefont {E.}~\bibnamefont
  {Cortese}}, \bibinfo {author} {\bibfnamefont {P.~G.}\ \bibnamefont
  {Lagoudakis}},\ and\ \bibinfo {author} {\bibfnamefont {S.}~\bibnamefont
  {De~Liberato}},\ }\href {https://doi.org/10.1103/PhysRevLett.119.043604}
  {\bibfield  {journal} {\bibinfo  {journal} {Physical Review Letters}\
  }\textbf {\bibinfo {volume} {119}},\ \bibinfo {pages} {043604} (\bibinfo
  {year} {2017})}\BibitemShut {NoStop}%
\bibitem [{\citenamefont {Phuc}(2022)}]{Phuc2022}%
  \BibitemOpen
  \bibfield  {author} {\bibinfo {author} {\bibfnamefont {N.~T.}\ \bibnamefont
  {Phuc}},\ }\href {https://doi.org/10.1063/5.0090463} {\bibfield  {journal}
  {\bibinfo  {journal} {The Journal of Chemical Physics}\ }\textbf {\bibinfo
  {volume} {156}},\ \bibinfo {pages} {234301} (\bibinfo {year}
  {2022})}\BibitemShut {NoStop}%
\bibitem [{\citenamefont {Englman}\ and\ \citenamefont
  {Jortner}(1970)}]{Englman1970}%
  \BibitemOpen
  \bibfield  {author} {\bibinfo {author} {\bibfnamefont {R.}~\bibnamefont
  {Englman}}\ and\ \bibinfo {author} {\bibfnamefont {J.}~\bibnamefont
  {Jortner}},\ }\href {https://doi.org/10.1080/00268977000100171} {\bibfield
  {journal} {\bibinfo  {journal} {Molecular Physics}\ }\textbf {\bibinfo
  {volume} {18}},\ \bibinfo {pages} {145} (\bibinfo {year} {1970})}\BibitemShut
  {NoStop}%
\bibitem [{\citenamefont {Fischer}(1970)}]{Fischer1970}%
  \BibitemOpen
  \bibfield  {author} {\bibinfo {author} {\bibfnamefont {S.}~\bibnamefont
  {Fischer}},\ }\href {https://doi.org/10.1063/1.1674470} {\bibfield  {journal}
  {\bibinfo  {journal} {The Journal of Chemical Physics}\ }\textbf {\bibinfo
  {volume} {53}},\ \bibinfo {pages} {3195} (\bibinfo {year}
  {1970})}\BibitemShut {NoStop}%
\bibitem [{\citenamefont {Jang}(2021)}]{Jang2021}%
  \BibitemOpen
  \bibfield  {author} {\bibinfo {author} {\bibfnamefont {S.~J.}\ \bibnamefont
  {Jang}},\ }\href {https://doi.org/10.1063/5.0068868} {\bibfield  {journal}
  {\bibinfo  {journal} {The Journal of Chemical Physics}\ }\textbf {\bibinfo
  {volume} {155}},\ \bibinfo {pages} {164106} (\bibinfo {year}
  {2021})}\BibitemShut {NoStop}%
\bibitem [{\citenamefont {Chynwat}\ and\ \citenamefont
  {Frank}(1995)}]{Chynwat1995}%
  \BibitemOpen
  \bibfield  {author} {\bibinfo {author} {\bibfnamefont {V.}~\bibnamefont
  {Chynwat}}\ and\ \bibinfo {author} {\bibfnamefont {H.~A.}\ \bibnamefont
  {Frank}},\ }\href {https://doi.org/10.1016/0301-0104(95)00017-I} {\bibfield
  {journal} {\bibinfo  {journal} {Chemical Physics}\ }\textbf {\bibinfo
  {volume} {194}},\ \bibinfo {pages} {237} (\bibinfo {year}
  {1995})}\BibitemShut {NoStop}%
\bibitem [{\citenamefont {Wilson}\ \emph {et~al.}(2001)\citenamefont {Wilson},
  \citenamefont {Chawdhury}, \citenamefont {Al-Mandhary}, \citenamefont
  {Younus}, \citenamefont {Khan}, \citenamefont {Raithby}, \citenamefont
  {Köhler},\ and\ \citenamefont {Friend}}]{Wilson2001}%
  \BibitemOpen
  \bibfield  {author} {\bibinfo {author} {\bibfnamefont {J.~S.}\ \bibnamefont
  {Wilson}}, \bibinfo {author} {\bibfnamefont {N.}~\bibnamefont {Chawdhury}},
  \bibinfo {author} {\bibfnamefont {M.~R.~A.}\ \bibnamefont {Al-Mandhary}},
  \bibinfo {author} {\bibfnamefont {M.}~\bibnamefont {Younus}}, \bibinfo
  {author} {\bibfnamefont {M.~S.}\ \bibnamefont {Khan}}, \bibinfo {author}
  {\bibfnamefont {P.~R.}\ \bibnamefont {Raithby}}, \bibinfo {author}
  {\bibfnamefont {A.}~\bibnamefont {Köhler}},\ and\ \bibinfo {author}
  {\bibfnamefont {R.~H.}\ \bibnamefont {Friend}},\ }\href
  {https://doi.org/10.1021/ja010986s} {\bibfield  {journal} {\bibinfo
  {journal} {Journal of the American Chemical Society}\ }\textbf {\bibinfo
  {volume} {123}},\ \bibinfo {pages} {9412} (\bibinfo {year}
  {2001})}\BibitemShut {NoStop}%
\bibitem [{\citenamefont {Caspar}\ \emph {et~al.}(1982)\citenamefont {Caspar},
  \citenamefont {Sullivan}, \citenamefont {Kober},\ and\ \citenamefont
  {Meyer}}]{Caspar1982}%
  \BibitemOpen
  \bibfield  {author} {\bibinfo {author} {\bibfnamefont {J.~V.}\ \bibnamefont
  {Caspar}}, \bibinfo {author} {\bibfnamefont {B.~P.}\ \bibnamefont
  {Sullivan}}, \bibinfo {author} {\bibfnamefont {E.~M.}\ \bibnamefont
  {Kober}},\ and\ \bibinfo {author} {\bibfnamefont {T.~J.}\ \bibnamefont
  {Meyer}},\ }\href {https://doi.org/10.1016/0009-2614(82)80114-0} {\bibfield
  {journal} {\bibinfo  {journal} {Chemical Physics Letters}\ }\textbf {\bibinfo
  {volume} {91}},\ \bibinfo {pages} {91} (\bibinfo {year} {1982})}\BibitemShut
  {NoStop}%
\bibitem [{\citenamefont {Martin}(1975)}]{Martin1975}%
  \BibitemOpen
  \bibfield  {author} {\bibinfo {author} {\bibfnamefont {M.~M.}\ \bibnamefont
  {Martin}},\ }\href {https://doi.org/10.1016/0009-2614(75)85598-9} {\bibfield
  {journal} {\bibinfo  {journal} {Chemical Physics Letters}\ }\textbf {\bibinfo
  {volume} {35}},\ \bibinfo {pages} {105} (\bibinfo {year} {1975})}\BibitemShut
  {NoStop}%
\bibitem [{\citenamefont {Tuong~Ly}\ \emph {et~al.}(2017)\citenamefont
  {Tuong~Ly}, \citenamefont {Chen-Cheng}, \citenamefont {Lin}, \citenamefont
  {Shiau}, \citenamefont {Liu}, \citenamefont {Chou}, \citenamefont {Tsao},
  \citenamefont {Huang},\ and\ \citenamefont {Chi}}]{TuongLy2017}%
  \BibitemOpen
  \bibfield  {author} {\bibinfo {author} {\bibfnamefont {K.}~\bibnamefont
  {Tuong~Ly}}, \bibinfo {author} {\bibfnamefont {R.-W.}\ \bibnamefont
  {Chen-Cheng}}, \bibinfo {author} {\bibfnamefont {H.-W.}\ \bibnamefont {Lin}},
  \bibinfo {author} {\bibfnamefont {Y.-J.}\ \bibnamefont {Shiau}}, \bibinfo
  {author} {\bibfnamefont {S.-H.}\ \bibnamefont {Liu}}, \bibinfo {author}
  {\bibfnamefont {P.-T.}\ \bibnamefont {Chou}}, \bibinfo {author}
  {\bibfnamefont {C.-S.}\ \bibnamefont {Tsao}}, \bibinfo {author}
  {\bibfnamefont {Y.-C.}\ \bibnamefont {Huang}},\ and\ \bibinfo {author}
  {\bibfnamefont {Y.}~\bibnamefont {Chi}},\ }\href
  {https://doi.org/10.1038/nphoton.2016.230} {\bibfield  {journal} {\bibinfo
  {journal} {Nature Photonics}\ }\textbf {\bibinfo {volume} {11}},\ \bibinfo
  {pages} {63} (\bibinfo {year} {2017})}\BibitemShut {NoStop}%
\bibitem [{\citenamefont {Wang}\ \emph {et~al.}(2017)\citenamefont {Wang},
  \citenamefont {Li}, \citenamefont {Gao}, \citenamefont {Wang}, \citenamefont
  {Zhang}, \citenamefont {Zhao}, \citenamefont {Lu}, \citenamefont {Yang},
  \citenamefont {Su},\ and\ \citenamefont {Ma}}]{Wang2017}%
  \BibitemOpen
  \bibfield  {author} {\bibinfo {author} {\bibfnamefont {C.}~\bibnamefont
  {Wang}}, \bibinfo {author} {\bibfnamefont {X.-L.}\ \bibnamefont {Li}},
  \bibinfo {author} {\bibfnamefont {Y.}~\bibnamefont {Gao}}, \bibinfo {author}
  {\bibfnamefont {L.}~\bibnamefont {Wang}}, \bibinfo {author} {\bibfnamefont
  {S.}~\bibnamefont {Zhang}}, \bibinfo {author} {\bibfnamefont
  {L.}~\bibnamefont {Zhao}}, \bibinfo {author} {\bibfnamefont {P.}~\bibnamefont
  {Lu}}, \bibinfo {author} {\bibfnamefont {B.}~\bibnamefont {Yang}}, \bibinfo
  {author} {\bibfnamefont {S.-J.}\ \bibnamefont {Su}},\ and\ \bibinfo {author}
  {\bibfnamefont {Y.}~\bibnamefont {Ma}},\ }\href
  {https://doi.org/10.1002/adom.201700441} {\bibfield  {journal} {\bibinfo
  {journal} {Advanced Optical Materials}\ }\textbf {\bibinfo {volume} {5}},\
  \bibinfo {pages} {1700441} (\bibinfo {year} {2017})}\BibitemShut {NoStop}%
\bibitem [{\citenamefont {Collado-Fregoso}\ \emph {et~al.}(2019)\citenamefont
  {Collado-Fregoso}, \citenamefont {Pugliese}, \citenamefont {Wojcik},
  \citenamefont {Benduhn}, \citenamefont {Bar-Or}, \citenamefont
  {Perdigón~Toro}, \citenamefont {Hörmann}, \citenamefont {Spoltore},
  \citenamefont {Vandewal}, \citenamefont {Hodgkiss},\ and\ \citenamefont
  {Neher}}]{ColladoFregoso2019}%
  \BibitemOpen
  \bibfield  {author} {\bibinfo {author} {\bibfnamefont {E.}~\bibnamefont
  {Collado-Fregoso}}, \bibinfo {author} {\bibfnamefont {S.~N.}\ \bibnamefont
  {Pugliese}}, \bibinfo {author} {\bibfnamefont {M.}~\bibnamefont {Wojcik}},
  \bibinfo {author} {\bibfnamefont {J.}~\bibnamefont {Benduhn}}, \bibinfo
  {author} {\bibfnamefont {E.}~\bibnamefont {Bar-Or}}, \bibinfo {author}
  {\bibfnamefont {L.}~\bibnamefont {Perdigón~Toro}}, \bibinfo {author}
  {\bibfnamefont {U.}~\bibnamefont {Hörmann}}, \bibinfo {author}
  {\bibfnamefont {D.}~\bibnamefont {Spoltore}}, \bibinfo {author}
  {\bibfnamefont {K.}~\bibnamefont {Vandewal}}, \bibinfo {author}
  {\bibfnamefont {J.~M.}\ \bibnamefont {Hodgkiss}},\ and\ \bibinfo {author}
  {\bibfnamefont {D.}~\bibnamefont {Neher}},\ }\href
  {https://doi.org/10.1021/jacs.8b09820} {\bibfield  {journal} {\bibinfo
  {journal} {Journal of the American Chemical Society}\ }\textbf {\bibinfo
  {volume} {141}},\ \bibinfo {pages} {2329} (\bibinfo {year}
  {2019})}\BibitemShut {NoStop}%
\bibitem [{\citenamefont {Wei}\ \emph {et~al.}(2020)\citenamefont {Wei},
  \citenamefont {Wang}, \citenamefont {Hu}, \citenamefont {Liao}, \citenamefont
  {Chen}, \citenamefont {Chang}, \citenamefont {Wang}, \citenamefont {Liu},
  \citenamefont {Chan}, \citenamefont {Liao}, \citenamefont {Hung},
  \citenamefont {Wang}, \citenamefont {Chen}, \citenamefont {Hsu},
  \citenamefont {Chi},\ and\ \citenamefont {Chou}}]{Wei2020}%
  \BibitemOpen
  \bibfield  {author} {\bibinfo {author} {\bibfnamefont {Y.-C.}\ \bibnamefont
  {Wei}}, \bibinfo {author} {\bibfnamefont {S.~F.}\ \bibnamefont {Wang}},
  \bibinfo {author} {\bibfnamefont {Y.}~\bibnamefont {Hu}}, \bibinfo {author}
  {\bibfnamefont {L.-S.}\ \bibnamefont {Liao}}, \bibinfo {author}
  {\bibfnamefont {D.-G.}\ \bibnamefont {Chen}}, \bibinfo {author}
  {\bibfnamefont {K.-H.}\ \bibnamefont {Chang}}, \bibinfo {author}
  {\bibfnamefont {C.-W.}\ \bibnamefont {Wang}}, \bibinfo {author}
  {\bibfnamefont {S.-H.}\ \bibnamefont {Liu}}, \bibinfo {author} {\bibfnamefont
  {W.-H.}\ \bibnamefont {Chan}}, \bibinfo {author} {\bibfnamefont {J.-L.}\
  \bibnamefont {Liao}}, \bibinfo {author} {\bibfnamefont {W.-Y.}\ \bibnamefont
  {Hung}}, \bibinfo {author} {\bibfnamefont {T.-H.}\ \bibnamefont {Wang}},
  \bibinfo {author} {\bibfnamefont {P.-T.}\ \bibnamefont {Chen}}, \bibinfo
  {author} {\bibfnamefont {H.-F.}\ \bibnamefont {Hsu}}, \bibinfo {author}
  {\bibfnamefont {Y.}~\bibnamefont {Chi}},\ and\ \bibinfo {author}
  {\bibfnamefont {P.-T.}\ \bibnamefont {Chou}},\ }\href
  {https://doi.org/10.1038/s41566-020-0653-6} {\bibfield  {journal} {\bibinfo
  {journal} {Nature Photonics}\ }\textbf {\bibinfo {volume} {14}},\ \bibinfo
  {pages} {570} (\bibinfo {year} {2020})}\BibitemShut {NoStop}%
\bibitem [{\citenamefont {Zhang}\ \emph {et~al.}(2021)\citenamefont {Zhang},
  \citenamefont {Zhang}, \citenamefont {Huang}, \citenamefont {Gillett},
  \citenamefont {Liu}, \citenamefont {Hu}, \citenamefont {Cui}, \citenamefont
  {Bin}, \citenamefont {Li}, \citenamefont {Wei},\ and\ \citenamefont
  {Duan}}]{Zhang2021}%
  \BibitemOpen
  \bibfield  {author} {\bibinfo {author} {\bibfnamefont {Y.}~\bibnamefont
  {Zhang}}, \bibinfo {author} {\bibfnamefont {D.}~\bibnamefont {Zhang}},
  \bibinfo {author} {\bibfnamefont {T.}~\bibnamefont {Huang}}, \bibinfo
  {author} {\bibfnamefont {A.~J.}\ \bibnamefont {Gillett}}, \bibinfo {author}
  {\bibfnamefont {Y.}~\bibnamefont {Liu}}, \bibinfo {author} {\bibfnamefont
  {D.}~\bibnamefont {Hu}}, \bibinfo {author} {\bibfnamefont {L.}~\bibnamefont
  {Cui}}, \bibinfo {author} {\bibfnamefont {Z.}~\bibnamefont {Bin}}, \bibinfo
  {author} {\bibfnamefont {G.}~\bibnamefont {Li}}, \bibinfo {author}
  {\bibfnamefont {J.}~\bibnamefont {Wei}},\ and\ \bibinfo {author}
  {\bibfnamefont {L.}~\bibnamefont {Duan}},\ }\href
  {https://doi.org/10.1002/anie.202107848} {\bibfield  {journal} {\bibinfo
  {journal} {Angewandte Chemie International Edition}\ }\textbf {\bibinfo
  {volume} {60}},\ \bibinfo {pages} {20498} (\bibinfo {year}
  {2021})}\BibitemShut {NoStop}%
\bibitem [{\citenamefont {Friedman}\ \emph {et~al.}(2021)\citenamefont
  {Friedman}, \citenamefont {Cosco}, \citenamefont {Atallah}, \citenamefont
  {Jia}, \citenamefont {Sletten},\ and\ \citenamefont {Caram}}]{Friedman2021}%
  \BibitemOpen
  \bibfield  {author} {\bibinfo {author} {\bibfnamefont {H.~C.}\ \bibnamefont
  {Friedman}}, \bibinfo {author} {\bibfnamefont {E.~D.}\ \bibnamefont {Cosco}},
  \bibinfo {author} {\bibfnamefont {T.~L.}\ \bibnamefont {Atallah}}, \bibinfo
  {author} {\bibfnamefont {S.}~\bibnamefont {Jia}}, \bibinfo {author}
  {\bibfnamefont {E.~M.}\ \bibnamefont {Sletten}},\ and\ \bibinfo {author}
  {\bibfnamefont {J.~R.}\ \bibnamefont {Caram}},\ }\href
  {https://doi.org/10.1016/j.chempr.2021.09.001} {\bibfield  {journal}
  {\bibinfo  {journal} {Chem}\ }\textbf {\bibinfo {volume} {7}},\ \bibinfo
  {pages} {3359} (\bibinfo {year} {2021})}\BibitemShut {NoStop}%
\bibitem [{\citenamefont {Byrne}\ \emph {et~al.}(1965)\citenamefont {Byrne},
  \citenamefont {McCoy},\ and\ \citenamefont {Ross}}]{Byrne1965}%
  \BibitemOpen
  \bibfield  {author} {\bibinfo {author} {\bibfnamefont {J.~P.}\ \bibnamefont
  {Byrne}}, \bibinfo {author} {\bibfnamefont {E.~F.}\ \bibnamefont {McCoy}},\
  and\ \bibinfo {author} {\bibfnamefont {I.~G.}\ \bibnamefont {Ross}},\ }\href
  {https://doi.org/10.1071/ch9651589} {\bibfield  {journal} {\bibinfo
  {journal} {Australian Journal of Chemistry}\ }\textbf {\bibinfo {volume}
  {18}},\ \bibinfo {pages} {1589} (\bibinfo {year} {1965})}\BibitemShut
  {NoStop}%
\bibitem [{\citenamefont {Strashko}\ and\ \citenamefont
  {Keeling}(2016)}]{Strashko2016}%
  \BibitemOpen
  \bibfield  {author} {\bibinfo {author} {\bibfnamefont {A.}~\bibnamefont
  {Strashko}}\ and\ \bibinfo {author} {\bibfnamefont {J.}~\bibnamefont
  {Keeling}},\ }\href {https://doi.org/10.1103/PhysRevA.94.023843} {\bibfield
  {journal} {\bibinfo  {journal} {Physical Review A}\ }\textbf {\bibinfo
  {volume} {94}},\ \bibinfo {pages} {023843} (\bibinfo {year}
  {2016})}\BibitemShut {NoStop}%
\bibitem [{\citenamefont {Heller}(2018)}]{Heller2018}%
  \BibitemOpen
  \bibfield  {author} {\bibinfo {author} {\bibfnamefont {E.~J.}\ \bibnamefont
  {Heller}},\ }\href
  {https://press.princeton.edu/books/hardcover/9780691163734/the-semiclassical-way-to-dynamics-and-spectroscopy}
  {\emph {\bibinfo {title} {The {Semiclassical} {Way} to {Dynamics} and
  {Spectroscopy}}}}\ (\bibinfo  {publisher} {Princeton University Press},\
  \bibinfo {year} {2018})\BibitemShut {NoStop}%
\bibitem [{\citenamefont {Marcus}(1956)}]{Marcus1956}%
  \BibitemOpen
  \bibfield  {author} {\bibinfo {author} {\bibfnamefont {R.~A.}\ \bibnamefont
  {Marcus}},\ }\href {https://doi.org/10.1063/1.1742723} {\bibfield  {journal}
  {\bibinfo  {journal} {The Journal of Chemical Physics}\ }\textbf {\bibinfo
  {volume} {24}},\ \bibinfo {pages} {966} (\bibinfo {year} {1956})}\BibitemShut
  {NoStop}%
\bibitem [{\citenamefont {Marcus}(1964)}]{Marcus1964}%
  \BibitemOpen
  \bibfield  {author} {\bibinfo {author} {\bibfnamefont {R.~A.}\ \bibnamefont
  {Marcus}},\ }\href {https://doi.org/10.1146/annurev.pc.15.100164.001103}
  {\bibfield  {journal} {\bibinfo  {journal} {Annual Review of Physical
  Chemistry}\ }\textbf {\bibinfo {volume} {15}},\ \bibinfo {pages} {155}
  (\bibinfo {year} {1964})}\BibitemShut {NoStop}%
\bibitem [{\citenamefont {Levich}(1966)}]{Levich1966}%
  \BibitemOpen
  \bibfield  {author} {\bibinfo {author} {\bibfnamefont {V.~G.}\ \bibnamefont
  {Levich}},\ }\href@noop {} {\bibfield  {journal} {\bibinfo  {journal} {Adv.
  Electrochem. Electrochem. Eng}\ }\textbf {\bibinfo {volume} {4}},\ \bibinfo
  {pages} {249} (\bibinfo {year} {1966})}\BibitemShut {NoStop}%
\bibitem [{\citenamefont {Jortner}(1976)}]{Jortner1976}%
  \BibitemOpen
  \bibfield  {author} {\bibinfo {author} {\bibfnamefont {J.}~\bibnamefont
  {Jortner}},\ }\href {https://doi.org/10.1063/1.432142} {\bibfield  {journal}
  {\bibinfo  {journal} {The Journal of Chemical Physics}\ }\textbf {\bibinfo
  {volume} {64}},\ \bibinfo {pages} {4860} (\bibinfo {year}
  {1976})}\BibitemShut {NoStop}%
\bibitem [{\citenamefont {Phuc}\ \emph {et~al.}(2020)\citenamefont {Phuc},
  \citenamefont {Trung},\ and\ \citenamefont {Ishizaki}}]{Phuc2020}%
  \BibitemOpen
  \bibfield  {author} {\bibinfo {author} {\bibfnamefont {N.~T.}\ \bibnamefont
  {Phuc}}, \bibinfo {author} {\bibfnamefont {P.~Q.}\ \bibnamefont {Trung}},\
  and\ \bibinfo {author} {\bibfnamefont {A.}~\bibnamefont {Ishizaki}},\ }\href
  {https://doi.org/10.1038/s41598-020-62899-8} {\bibfield  {journal} {\bibinfo
  {journal} {Scientific Reports}\ }\textbf {\bibinfo {volume} {10}},\ \bibinfo
  {pages} {7318} (\bibinfo {year} {2020})}\BibitemShut {NoStop}%
\bibitem [{\citenamefont {Nitzan}(2006)}]{Nitzan2006-EGL}%
  \BibitemOpen
  \bibfield  {author} {\bibinfo {author} {\bibfnamefont {A.}~\bibnamefont
  {Nitzan}},\ }in\ \href@noop {} {\emph {\bibinfo {booktitle} {Chemical
  Dynamics in Condensed Phases: Relaxation, Transfer and Reactions in Condensed
  Molecular Systems}}}\ (\bibinfo  {publisher} {Oxford University Press},\
  \bibinfo {year} {2006})\ Chap.\ \bibinfo {chapter} {12.5}, pp.\ \bibinfo
  {pages} {439--449}\BibitemShut {NoStop}%
\bibitem [{\citenamefont {Morse}\ and\ \citenamefont
  {Feshbach}(1953)}]{Morse1953}%
  \BibitemOpen
  \bibfield  {author} {\bibinfo {author} {\bibfnamefont {P.~M.}\ \bibnamefont
  {Morse}}\ and\ \bibinfo {author} {\bibfnamefont {H.}~\bibnamefont
  {Feshbach}},\ }\href
  {https://www.abebooks.com/9780070433168/Methods-Theoretical-Physics-Part-International-007043316X/plp}
  {\emph {\bibinfo {title} {Methods of {Theoretical} {Physics}}}},\ {Part} {I}\
  (\bibinfo  {publisher} {McGraw-Hill Book Company},\ \bibinfo {year}
  {1953})\BibitemShut {NoStop}%
\bibitem [{\citenamefont {Kessing}\ \emph {et~al.}(2022)\citenamefont
  {Kessing}, \citenamefont {Yang}, \citenamefont {Manmana},\ and\ \citenamefont
  {Cao}}]{Kessing2022}%
  \BibitemOpen
  \bibfield  {author} {\bibinfo {author} {\bibfnamefont {R.~K.}\ \bibnamefont
  {Kessing}}, \bibinfo {author} {\bibfnamefont {P.-Y.}\ \bibnamefont {Yang}},
  \bibinfo {author} {\bibfnamefont {S.~R.}\ \bibnamefont {Manmana}},\ and\
  \bibinfo {author} {\bibfnamefont {J.}~\bibnamefont {Cao}},\ }\href
  {https://doi.org/10.1021/acs.jpclett.2c01455} {\bibfield  {journal} {\bibinfo
   {journal} {The Journal of Physical Chemistry Letters}\ }\textbf {\bibinfo
  {volume} {13}},\ \bibinfo {pages} {6831} (\bibinfo {year}
  {2022})}\BibitemShut {NoStop}%
\bibitem [{\citenamefont {Cao}(2022)}]{Cao2022}%
  \BibitemOpen
  \bibfield  {author} {\bibinfo {author} {\bibfnamefont {J.}~\bibnamefont
  {Cao}},\ }\href {https://doi.org/10.1021/acs.jpclett.2c02707} {\bibfield
  {journal} {\bibinfo  {journal} {The Journal of Physical Chemistry Letters}\
  ,\ \bibinfo {pages} {10943}} (\bibinfo {year} {2022})}\BibitemShut {NoStop}%
\bibitem [{\citenamefont {Yang}\ and\ \citenamefont {Cao}(2021)}]{Yang2021}%
  \BibitemOpen
  \bibfield  {author} {\bibinfo {author} {\bibfnamefont {P.-Y.}\ \bibnamefont
  {Yang}}\ and\ \bibinfo {author} {\bibfnamefont {J.}~\bibnamefont {Cao}},\
  }\href {https://doi.org/10.1021/acs.jpclett.1c02210} {\bibfield  {journal}
  {\bibinfo  {journal} {The Journal of Physical Chemistry Letters}\ }\textbf
  {\bibinfo {volume} {12}},\ \bibinfo {pages} {9531} (\bibinfo {year}
  {2021})}\BibitemShut {NoStop}%
\bibitem [{\citenamefont {Kansanen}(2022)}]{Kansanen2022}%
  \BibitemOpen
  \bibfield  {author} {\bibinfo {author} {\bibfnamefont {K.~S.~U.}\
  \bibnamefont {Kansanen}},\ }\href {https://doi.org/10.48550/ARXIV.2204.13490}
  {\bibinfo {title} {Theory for polaritonic quantum tunneling}} (\bibinfo
  {year} {2022})\BibitemShut {NoStop}%
\bibitem [{\citenamefont {Du}\ \emph {et~al.}(2022)\citenamefont {Du},
  \citenamefont {Poh},\ and\ \citenamefont {Yuen-Zhou}}]{Du2022-PGH}%
  \BibitemOpen
  \bibfield  {author} {\bibinfo {author} {\bibfnamefont {M.}~\bibnamefont
  {Du}}, \bibinfo {author} {\bibfnamefont {Y.~R.}\ \bibnamefont {Poh}},\ and\
  \bibinfo {author} {\bibfnamefont {J.}~\bibnamefont {Yuen-Zhou}},\ }\href
  {https://doi.org/10.48550/ARXIV.2211.05820} {\bibinfo {title}
  {Vibropolaritonic {Reaction} {Rates} in the {Collective} {Strong} {Coupling}
  {Regime}: {Pollak}-{Grabert}-{Hänggi} {Theory}}} (\bibinfo {year}
  {2022})\BibitemShut {NoStop}%
\bibitem [{\citenamefont {Martínez-Martínez}\ \emph
  {et~al.}(2019)\citenamefont {Martínez-Martínez}, \citenamefont {Eizner},
  \citenamefont {Kéna-Cohen},\ and\ \citenamefont
  {Yuen-Zhou}}]{MartinezMartinez2019}%
  \BibitemOpen
  \bibfield  {author} {\bibinfo {author} {\bibfnamefont {L.~A.}\ \bibnamefont
  {Martínez-Martínez}}, \bibinfo {author} {\bibfnamefont {E.}~\bibnamefont
  {Eizner}}, \bibinfo {author} {\bibfnamefont {S.}~\bibnamefont
  {Kéna-Cohen}},\ and\ \bibinfo {author} {\bibfnamefont {J.}~\bibnamefont
  {Yuen-Zhou}},\ }\href {https://doi.org/10.1063/1.5100192} {\bibfield
  {journal} {\bibinfo  {journal} {The Journal of Chemical Physics}\ }\textbf
  {\bibinfo {volume} {151}},\ \bibinfo {pages} {054106} (\bibinfo {year}
  {2019})}\BibitemShut {NoStop}%
\bibitem [{\citenamefont {Maciejewski}\ \emph {et~al.}(1984)\citenamefont
  {Maciejewski}, \citenamefont {Safarzadeh-Amiri}, \citenamefont {Verrall},\
  and\ \citenamefont {Steer}}]{Maciejewski1984}%
  \BibitemOpen
  \bibfield  {author} {\bibinfo {author} {\bibfnamefont {A.}~\bibnamefont
  {Maciejewski}}, \bibinfo {author} {\bibfnamefont {A.}~\bibnamefont
  {Safarzadeh-Amiri}}, \bibinfo {author} {\bibfnamefont {R.~E.}\ \bibnamefont
  {Verrall}},\ and\ \bibinfo {author} {\bibfnamefont {R.~P.}\ \bibnamefont
  {Steer}},\ }\href {https://doi.org/10.1016/0301-0104(84)85054-5} {\bibfield
  {journal} {\bibinfo  {journal} {Chemical Physics}\ }\textbf {\bibinfo
  {volume} {87}},\ \bibinfo {pages} {295} (\bibinfo {year} {1984})}\BibitemShut
  {NoStop}%
\bibitem [{\citenamefont {Chikkaraddy}\ \emph {et~al.}(2016)\citenamefont
  {Chikkaraddy}, \citenamefont {de~Nijs}, \citenamefont {Benz}, \citenamefont
  {Barrow}, \citenamefont {Scherman}, \citenamefont {Rosta}, \citenamefont
  {Demetriadou}, \citenamefont {Fox}, \citenamefont {Hess},\ and\ \citenamefont
  {Baumberg}}]{Chikkaraddy2016}%
  \BibitemOpen
  \bibfield  {author} {\bibinfo {author} {\bibfnamefont {R.}~\bibnamefont
  {Chikkaraddy}}, \bibinfo {author} {\bibfnamefont {B.}~\bibnamefont
  {de~Nijs}}, \bibinfo {author} {\bibfnamefont {F.}~\bibnamefont {Benz}},
  \bibinfo {author} {\bibfnamefont {S.~J.}\ \bibnamefont {Barrow}}, \bibinfo
  {author} {\bibfnamefont {O.~A.}\ \bibnamefont {Scherman}}, \bibinfo {author}
  {\bibfnamefont {E.}~\bibnamefont {Rosta}}, \bibinfo {author} {\bibfnamefont
  {A.}~\bibnamefont {Demetriadou}}, \bibinfo {author} {\bibfnamefont
  {P.}~\bibnamefont {Fox}}, \bibinfo {author} {\bibfnamefont {O.}~\bibnamefont
  {Hess}},\ and\ \bibinfo {author} {\bibfnamefont {J.~J.}\ \bibnamefont
  {Baumberg}},\ }\href {https://doi.org/10.1038/nature17974} {\bibfield
  {journal} {\bibinfo  {journal} {Nature}\ }\textbf {\bibinfo {volume} {535}},\
  \bibinfo {pages} {127} (\bibinfo {year} {2016})}\BibitemShut {NoStop}%
\bibitem [{\citenamefont {Bitton}\ and\ \citenamefont
  {Haran}(2022)}]{Bitton2022}%
  \BibitemOpen
  \bibfield  {author} {\bibinfo {author} {\bibfnamefont {O.}~\bibnamefont
  {Bitton}}\ and\ \bibinfo {author} {\bibfnamefont {G.}~\bibnamefont {Haran}},\
  }\href {https://doi.org/10.1021/acs.accounts.2c00028} {\bibfield  {journal}
  {\bibinfo  {journal} {Accounts of Chemical Research}\ }\textbf {\bibinfo
  {volume} {55}},\ \bibinfo {pages} {1659} (\bibinfo {year}
  {2022})}\BibitemShut {NoStop}%
\bibitem [{\citenamefont {Ansari}\ \emph {et~al.}(2022)\citenamefont {Ansari},
  \citenamefont {Heller}, \citenamefont {Trenins},\ and\ \citenamefont
  {Richardson}}]{Ansari2022}%
  \BibitemOpen
  \bibfield  {author} {\bibinfo {author} {\bibfnamefont {I.~M.}\ \bibnamefont
  {Ansari}}, \bibinfo {author} {\bibfnamefont {E.~R.}\ \bibnamefont {Heller}},
  \bibinfo {author} {\bibfnamefont {G.}~\bibnamefont {Trenins}},\ and\ \bibinfo
  {author} {\bibfnamefont {J.~O.}\ \bibnamefont {Richardson}},\ }\href
  {https://doi.org/10.1098/rsta.2020.0378} {\bibfield  {journal} {\bibinfo
  {journal} {Philosophical Transactions of the Royal Society A: Mathematical,
  Physical and Engineering Sciences}\ }\textbf {\bibinfo {volume} {380}},\
  \bibinfo {pages} {20200378} (\bibinfo {year} {2022})}\BibitemShut {NoStop}%
\bibitem [{\citenamefont {Fang}\ \emph {et~al.}(2019)\citenamefont {Fang},
  \citenamefont {Thapa},\ and\ \citenamefont {Richardson}}]{Fang2019}%
  \BibitemOpen
  \bibfield  {author} {\bibinfo {author} {\bibfnamefont {W.}~\bibnamefont
  {Fang}}, \bibinfo {author} {\bibfnamefont {M.~J.}\ \bibnamefont {Thapa}},\
  and\ \bibinfo {author} {\bibfnamefont {J.~O.}\ \bibnamefont {Richardson}},\
  }\href {https://doi.org/10.1063/1.5131092} {\bibfield  {journal} {\bibinfo
  {journal} {The Journal of Chemical Physics}\ }\textbf {\bibinfo {volume}
  {151}},\ \bibinfo {pages} {214101} (\bibinfo {year} {2019})}\BibitemShut
  {NoStop}%
\bibitem [{\citenamefont {Li}\ \emph {et~al.}(2022)\citenamefont {Li},
  \citenamefont {Nitzan}, \citenamefont {Hammes-Schiffer},\ and\ \citenamefont
  {Subotnik}}]{Li2022}%
  \BibitemOpen
  \bibfield  {author} {\bibinfo {author} {\bibfnamefont {T.~E.}\ \bibnamefont
  {Li}}, \bibinfo {author} {\bibfnamefont {A.}~\bibnamefont {Nitzan}}, \bibinfo
  {author} {\bibfnamefont {S.}~\bibnamefont {Hammes-Schiffer}},\ and\ \bibinfo
  {author} {\bibfnamefont {J.~E.}\ \bibnamefont {Subotnik}},\ }\href
  {https://doi.org/10.1021/acs.jpclett.2c00613} {\bibfield  {journal} {\bibinfo
   {journal} {The Journal of Physical Chemistry Letters}\ }\textbf {\bibinfo
  {volume} {13}},\ \bibinfo {pages} {3890} (\bibinfo {year}
  {2022})}\BibitemShut {NoStop}%
\bibitem [{\citenamefont {Rokaj}\ \emph {et~al.}(2018)\citenamefont {Rokaj},
  \citenamefont {Welakuh}, \citenamefont {Ruggenthaler},\ and\ \citenamefont
  {Rubio}}]{Rokaj2018}%
  \BibitemOpen
  \bibfield  {author} {\bibinfo {author} {\bibfnamefont {V.}~\bibnamefont
  {Rokaj}}, \bibinfo {author} {\bibfnamefont {D.~M.}\ \bibnamefont {Welakuh}},
  \bibinfo {author} {\bibfnamefont {M.}~\bibnamefont {Ruggenthaler}},\ and\
  \bibinfo {author} {\bibfnamefont {A.}~\bibnamefont {Rubio}},\ }\href
  {https://doi.org/10.1088/1361-6455/aa9c99} {\bibfield  {journal} {\bibinfo
  {journal} {Journal of Physics B: Atomic, Molecular and Optical Physics}\
  }\textbf {\bibinfo {volume} {51}},\ \bibinfo {pages} {034005} (\bibinfo
  {year} {2018})}\BibitemShut {NoStop}%
\bibitem [{\citenamefont {Mandal}\ \emph {et~al.}(2020)\citenamefont {Mandal},
  \citenamefont {Krauss},\ and\ \citenamefont {Huo}}]{Mandal2020}%
  \BibitemOpen
  \bibfield  {author} {\bibinfo {author} {\bibfnamefont {A.}~\bibnamefont
  {Mandal}}, \bibinfo {author} {\bibfnamefont {T.~D.}\ \bibnamefont {Krauss}},\
  and\ \bibinfo {author} {\bibfnamefont {P.}~\bibnamefont {Huo}},\ }\href
  {https://doi.org/10.1021/acs.jpcb.0c03227} {\bibfield  {journal} {\bibinfo
  {journal} {The Journal of Physical Chemistry B}\ }\textbf {\bibinfo {volume}
  {124}},\ \bibinfo {pages} {6321} (\bibinfo {year} {2020})}\BibitemShut
  {NoStop}%
\bibitem [{\citenamefont {Cohen-Tannoudji}\ \emph {et~al.}(1989)\citenamefont
  {Cohen-Tannoudji}, \citenamefont {Dupont-Roc},\ and\ \citenamefont
  {Grynberg}}]{CohenTannoudji1989}%
  \BibitemOpen
  \bibfield  {author} {\bibinfo {author} {\bibfnamefont {C.}~\bibnamefont
  {Cohen-Tannoudji}}, \bibinfo {author} {\bibfnamefont {J.}~\bibnamefont
  {Dupont-Roc}},\ and\ \bibinfo {author} {\bibfnamefont {G.}~\bibnamefont
  {Grynberg}},\ }\href@noop {} {\emph {\bibinfo {title} {{Photons} and
  {Atoms}}}}\ (\bibinfo  {publisher} {Wiley-VCH},\ \bibinfo {year}
  {1989})\BibitemShut {NoStop}%
\bibitem [{\citenamefont {Rzażewski}\ \emph {et~al.}(1975)\citenamefont
  {Rzażewski}, \citenamefont {Wódkiewicz},\ and\ \citenamefont
  {Żakowicz}}]{Rzazewski1975}%
  \BibitemOpen
  \bibfield  {author} {\bibinfo {author} {\bibfnamefont {K.}~\bibnamefont
  {Rzażewski}}, \bibinfo {author} {\bibfnamefont {K.}~\bibnamefont
  {Wódkiewicz}},\ and\ \bibinfo {author} {\bibfnamefont {W.}~\bibnamefont
  {Żakowicz}},\ }\href {https://doi.org/10.1103/PhysRevLett.35.432} {\bibfield
   {journal} {\bibinfo  {journal} {Physical Review Letters}\ }\textbf {\bibinfo
  {volume} {35}},\ \bibinfo {pages} {432} (\bibinfo {year} {1975})}\BibitemShut
  {NoStop}%
\bibitem [{\citenamefont {Hopfield}(1958)}]{Hopfield1958}%
  \BibitemOpen
  \bibfield  {author} {\bibinfo {author} {\bibfnamefont {J.~J.}\ \bibnamefont
  {Hopfield}},\ }\href {https://doi.org/10.1103/PhysRev.112.1555} {\bibfield
  {journal} {\bibinfo  {journal} {Physical Review}\ }\textbf {\bibinfo {volume}
  {112}},\ \bibinfo {pages} {1555} (\bibinfo {year} {1958})}\BibitemShut
  {NoStop}%
\bibitem [{\citenamefont {Ribeiro}\ \emph
  {et~al.}(2018{\natexlab{a}})\citenamefont {Ribeiro}, \citenamefont
  {Martínez-Martínez}, \citenamefont {Du}, \citenamefont
  {Campos-Gonzalez-Angulo},\ and\ \citenamefont {Yuen-Zhou}}]{Ribeiro2018}%
  \BibitemOpen
  \bibfield  {author} {\bibinfo {author} {\bibfnamefont {R.~F.}\ \bibnamefont
  {Ribeiro}}, \bibinfo {author} {\bibfnamefont {L.~A.}\ \bibnamefont
  {Martínez-Martínez}}, \bibinfo {author} {\bibfnamefont {M.}~\bibnamefont
  {Du}}, \bibinfo {author} {\bibfnamefont {J.}~\bibnamefont
  {Campos-Gonzalez-Angulo}},\ and\ \bibinfo {author} {\bibfnamefont
  {J.}~\bibnamefont {Yuen-Zhou}},\ }\href {https://doi.org/10.1039/C8SC01043A}
  {\bibfield  {journal} {\bibinfo  {journal} {Chemical Science}\ }\textbf
  {\bibinfo {volume} {9}},\ \bibinfo {pages} {6325} (\bibinfo {year}
  {2018}{\natexlab{a}})}\BibitemShut {NoStop}%
\bibitem [{\citenamefont {Proukakis}\ \emph {et~al.}(2017)\citenamefont
  {Proukakis}, \citenamefont {Snoke},\ and\ \citenamefont
  {Littlewood}}]{Proukakis2017}%
  \BibitemOpen
  \bibfield  {author} {\bibinfo {author} {\bibfnamefont {N.~P.}\ \bibnamefont
  {Proukakis}}, \bibinfo {author} {\bibfnamefont {D.~W.}\ \bibnamefont
  {Snoke}},\ and\ \bibinfo {author} {\bibfnamefont {P.~B.}\ \bibnamefont
  {Littlewood}},\ }\href@noop {} {\emph {\bibinfo {title} {Universal {Themes}
  of {Bose}–{Einstein} {Condensation}}}}\ (\bibinfo  {publisher} {Cambridge
  University Press},\ \bibinfo {year} {2017})\BibitemShut {NoStop}%
\bibitem [{\citenamefont {Palma}\ and\ \citenamefont
  {Morales}(1983)}]{Palma1983}%
  \BibitemOpen
  \bibfield  {author} {\bibinfo {author} {\bibfnamefont {A.}~\bibnamefont
  {Palma}}\ and\ \bibinfo {author} {\bibfnamefont {J.}~\bibnamefont
  {Morales}},\ }\href@noop {} {\bibfield  {journal} {\bibinfo  {journal}
  {International Journal of Quantum Chemistry}\ }\textbf {\bibinfo {volume}
  {24}},\ \bibinfo {pages} {393} (\bibinfo {year} {1983})}\BibitemShut
  {NoStop}%
\bibitem [{\citenamefont {Ribeiro}\ \emph
  {et~al.}(2018{\natexlab{b}})\citenamefont {Ribeiro}, \citenamefont
  {Dunkelberger}, \citenamefont {Xiang}, \citenamefont {Xiong}, \citenamefont
  {Simpkins}, \citenamefont {Owrutsky},\ and\ \citenamefont
  {Yuen-Zhou}}]{Ribeiro2018-pump}%
  \BibitemOpen
  \bibfield  {author} {\bibinfo {author} {\bibfnamefont {R.~F.}\ \bibnamefont
  {Ribeiro}}, \bibinfo {author} {\bibfnamefont {A.~D.}\ \bibnamefont
  {Dunkelberger}}, \bibinfo {author} {\bibfnamefont {B.}~\bibnamefont {Xiang}},
  \bibinfo {author} {\bibfnamefont {W.}~\bibnamefont {Xiong}}, \bibinfo
  {author} {\bibfnamefont {B.~S.}\ \bibnamefont {Simpkins}}, \bibinfo {author}
  {\bibfnamefont {J.~C.}\ \bibnamefont {Owrutsky}},\ and\ \bibinfo {author}
  {\bibfnamefont {J.}~\bibnamefont {Yuen-Zhou}},\ }\href
  {https://doi.org/10.1021/acs.jpclett.8b01176} {\bibfield  {journal} {\bibinfo
   {journal} {The Journal of Physical Chemistry Letters}\ }\textbf {\bibinfo
  {volume} {9}},\ \bibinfo {pages} {3766} (\bibinfo {year}
  {2018}{\natexlab{b}})}\BibitemShut {NoStop}%
\end{thebibliography}
\end{document}